\documentclass[%
 aip,
pof,
 amsmath,amssymb,
preprint,%
]{revtex4-1}

\usepackage {CJK}

\usepackage{graphicx}% Include figure files
\usepackage{dcolumn}% Align table columns on decimal point
\usepackage{bm}% bold math
\usepackage[utf8]{inputenc}
\usepackage[T1]{fontenc}
\usepackage{mathptmx}

\usepackage{epstopdf}

\begin{document}

\begin{CJK*}{UTF8}{} % Use default fonts from CJK (see below)

\preprint{Preprint submit to \emph{Physics of Fluids}}

\title{Correlation of internal flow structure with heat transfer efficiency in turbulent Rayleigh-B\'enard convection}
% Force line breaks with \\

\author{Ao Xu}
\email{Corresponding author: axu@nwpu.edu.cn (Ao Xu)}
 \affiliation{School of Aeronautics, Northwestern Polytechnical University, Xi'an 710072, China}%

\author{Xin Chen}
 \affiliation{School of Aeronautics, Northwestern Polytechnical University, Xi'an 710072, China}%

\author{Feng Wang}
 \affiliation{School of Aeronautics, Northwestern Polytechnical University, Xi'an 710072, China}%

\author{Heng-Dong Xi}
\email{Corresponding author: hengdongxi@nwpu.edu.cn (Heng-Dong Xi)}
 \affiliation{School of Aeronautics, Northwestern Polytechnical University, Xi'an 710072, China}%

\date{\today}% It is always \today, today,

\begin{abstract}
To understand how internal flow structures manifest themselves in the global heat transfer, we study the correlation between different flow modes and the instantaneous Nusselt number ($Nu$) in a two-dimensional square Rayleigh-B\'enard convection cell.
High-resolution and long-time direct numerical simulations are carried out for Rayleigh numbers between $10^{7}$ and $10^{9}$ and a Prandtl number of 5.3.
The investigated Nusselt numbers include the volume-averaged $Nu_{\text{vol}}$, the wall-averaged $Nu_{\text{wall}}$, the kinetic energy dissipation based $Nu_{\text{kinetic}}$, and the thermal energy dissipation based $Nu_{\text{thermal}}$.
The Fourier mode decomposition and proper orthogonal decomposition are adopted to extract the coherent flow structure.
Our results show that the single-roll mode, the horizontally stacked double-roll mode, and the quadrupolar flow mode are more efficient for heat transfer on average.
In contrast, the vertically stacked double-roll mode is inefficient for heat transfer on average.
The volume-averaged $Nu_{\text{vol}}$ and the kinetic energy dissipation based $Nu_{\text{kinetic}}$ can better reproduce the correlation of internal flow structures with heat transfer efficiency than that of the wall-averaged $Nu_{\text{wall}}$ and the thermal energy dissipation based $Nu_{\text{thermal}}$, even though these four Nusselt numbers give consistent time-averaged mean values.
The ensemble-averaged time trace of $Nu$ during flow reversal shows that only the volume-averaged $Nu_{\text{vol}}$ can reproduce the overshoot phenomena that is observed in the previous experimental study.
Our results reveal that the proper choice of $Nu$ is critical to obtain a meaningful interpretation.
\footnote{
%%The following article has been submitted to Physics of Fluids.
%%After it is published, it will be found at Link (https://publishing.aip.org/resources/librarians/products/journals/).
This article may be downloaded for personal use only.
Any other use requires prior permission of the author and AIP Publishing.
This article appeared in Xu et al., Phys. Fluids \textbf{32}, 105112 (2020) and may be found at \url{https://doi.org/10.1063/5.0024408}.
}
\end{abstract}

\maketitle
\end{CJK*}

\section{\label{Section1}Introduction}

Thermal convection occurs ubiquitously in nature and has wide applications in industry.
A paradigm for the study of thermal convection is the Rayleigh-B\'enard (RB) convection, which is a fluid layer heated from the bottom and cooled from the top \cite{ahlers2009heat,lohse2010small,chilla2012new,xia2013current,mazzino2017two,wang2020vibration}.
The control parameters of the RB system include the Rayleigh number ($Ra$) and the Prandtl number ($Pr$).
$Ra$ describes the strength of the buoyancy force relative to thermal and viscous dissipative effects as $Ra = \beta g \Delta_{T}H^{3}/(\nu \kappa)$.
$Pr$ represents the thermophysical fluid properties as $Pr = \nu/\kappa$.
Here, $H$ is the fluid layer height and $\Delta_{T}$ is the imposed temperature difference. $\beta$, $\kappa$, and $\nu$ are the thermal expansion coefficient, thermal diffusivity, and kinematic viscosity of the fluid, respectively.
$g$ is the gravitational acceleration.
One of the response parameters of the RB system is the Nusselt number ($Nu$), which describes the global heat transfer efficiency of the system and is generally calculated as $Nu = J_{\text{total}}/J_{\text{conduction}}$.
Here, $J_{\text{total}}$ is the total heat flux and $J_{\text{conduction}}$ is the heat flux due to pure conduction across the bottom and top walls.
In an RB experiment \cite{heslot1987transitions,castaing1989scaling,xia1997turbulent,shang2008scaling,yang2020experimental}, the instantaneous $Nu$ is calculated as $Nu(t) = (Q/S)/[\chi \Delta {T}(t)/H]$, where $Q$ is the power supplied to the RB convection cell, $S$ is the cross-sectional area of the cell, and $\chi$ is the thermal conductivity of the fluid.
The temperature difference between the bottom and top walls is $\Delta T(t) = T_{b}(t)-T_{t}(t)$, where the temperatures of the bottom and top walls $T_{b}(t)$ and $T_{t}(t)$ are based on the average values of the two embedded thermistors in each plate.

In an RB direct numerical simulation (DNS), there are several different approaches to calculate $Nu$.
Before briefly reviewing these approaches, we first introduce the following dimensionless variables:
\begin{equation}
    \begin{split}
& \mathbf{x}/H \rightarrow \mathbf{x}^{*}, \ \ \ \
  t/ \sqrt{H/ (\beta g \Delta_{T})}  \rightarrow t^{*}, \ \ \ \
  \mathbf{u}/ \sqrt{\beta g H \Delta_{T}}  \rightarrow \mathbf{u}^{*}, \ \ \ \ \\
& p/\left( \rho_{0}g\beta \Delta_{T} L_{0} \right) \rightarrow p^{*}, \ \ \ \
  \left( T-T_{0}\right)/\Delta_{T} \rightarrow T^{*} \ \ \ \
    \end{split}
\end{equation}
Here, $T_{0}$ denotes the reference temperature.
$\mathbf{x}$, $t$, $\mathbf{u}$, $p$, and $T$ are the position, time, velocity, pressure, and temperature, respectively.
Their counterparts with the asterisk superscript ($^{*}$) denote the dimensionless variables.
The first approach to calculate $Nu$ is based on volume-averaged velocity and temperature fields  \cite{kerr1996rayleigh,verzicco2003numerical} as $Nu_{\text{vol}} =\sqrt{RaPr} \langle w^{*} T^{*} \rangle_{V,t}+1$, where $w^{*}$ is the dimensionless vertical velocity component.
Here, $\langle \cdots \rangle_{V,t}$ denotes the ensemble average over the whole convection cell and over the time.
We assume the two parallel hot and cold walls are perpendicular to the vertical direction $z$.
In this approach, the volume-averaged heat flux across the cell is $\langle wT- \kappa \partial  T/\partial z \rangle_{V,t}$, and the heat flux due to pure conduction is $\kappa \Delta_{T}/H$; thus, $Nu_{\text{vol}} = \langle wT-k \partial T/\partial z \rangle_{V,t}/(\kappa \Delta_{T}/H)= \sqrt{RaPr} \langle w^{*} T^{*} \rangle_{V,t}-\langle \partial T^{*}/\partial z^{*} \rangle_{V,t}$;
meanwhile, the top and the bottom walls remain at constant cold and hot temperatures, respectively, and we then have the term $\langle \partial T^{*}/\partial z^{*} \rangle_{V,t}=-1$.
%% thus, $Nu_{\text{vol}} = \sqrt{RaPr} \langle w^{*} T^{*} \rangle_{V,t}+1$.
The second approach is to directly calculate the mean heat flux at the top and bottom walls\cite{kerr1996rayleigh,verzicco2003numerical} as $Nu_{\text{wall}} = -(\langle \partial_z^{*} T^{*} \rangle_{\text{top},t}+ \langle \partial_z^{*} T^{*} \rangle_{\text{bottom},t})/2$.
Here, $\langle \cdots \rangle_{\text{top}/\text{bottom},t}$ denotes the ensemble average over the top (or bottom) wall and over the time.
This approach takes advantage of the no-slip boundary conditions at the walls, thus $Nu_{\text{top}/\text{bottom}} =\langle -\kappa \partial T/ \partial z \rangle_{\text{top}/\text{bottom},t}/( \kappa \Delta_{T}/H)= -\langle \partial T^{*}/\partial z^{*} \rangle_{\text{top} / \text{bottom},t}$, and we then take the mean value of $Nu_{\text{top}}$ and $Nu_{\text{bottom}}$ as $Nu_{\text{wall}}$.
The third and fourth approaches are based on kinetic and thermal energy dissipation fields as $Nu_{\text{kinetic}}=1+\sqrt{RaPr}\langle \varepsilon_{u}^{*} \rangle_{V,t}$ and $Nu_{\text{thermal}}=\sqrt{Ra Pr}\langle \varepsilon _{T}^{*} \rangle_{V,t}$, respectively.
Here, the kinetic and thermal energy dissipation rates in the dimensional form are defined as  $\varepsilon_{u}(\mathbf{x},t)=(\nu/2)\sum_{ij}[\partial_{i}u_{j}(\mathbf{x},t)+\partial_{j}u_{i}(\mathbf{x},t)]^{2}$ and $\varepsilon_{T}(\mathbf{x},t)=\kappa\sum_{i}[\partial_{i}T(\mathbf{x},t)]^{2}$, respectively.
These two approaches utilize the exact relations of Nusselt numbers and global averages of the kinetic and thermal energy dissipation \cite{shraiman1990heat,siggia1994high} as $\langle \varepsilon_{u}^{*} \rangle_{V,t}=(Nu-1)/\sqrt{RaPr}$ and  $\langle \varepsilon_{T}^{*} \rangle_{V,t}=Nu/\sqrt{RaPr}$, respectively.
These exact relations were obtained by averaging the equation of motion and heat equation, which further form the backbone of the Grossmann-Lohse (GL) theory on turbulent heat transfer \cite{grossmann2000scaling,grossmann2004fluctuations}.
It should be noted that the above four approaches to calculate the Nusselt numbers would give consistent values if the DNS is well resolved and statistically convergent, but not \emph{vice versa}.
For example, Kooij et al. \cite{kooij2018comparison} observed ripples in instantaneous snapshots of the temperature field near sharp gradients when the simulation is under-resolved, while the above four Nusselt numbers from the simulation still look reasonable.

Previous studies have shown connections between $Nu$ and the flow structures in the RB system \cite{sun2005azimuthal,xi2008flow,weiss2011turbulent,van2011connecting,van2012flow,xi2016higher}.
Sun et al. \cite{sun2005azimuthal} experimentally measured $Nu$ in a cylindrical  leveled  cell (in which the  large-scale circulation plane azimuthal sweeps) and in a tilted cell (in which the large-scale circulation, i.e., the LSC, is locked in a particular orientation).
Results showed that $Nu$ is larger in the leveled cell than that in the tilted one, thus demonstrating that different flow structures can give rise to different values of $Nu$.
Xi and Xia \cite{xi2008flow} further observed both the single-roll structure and the double-roll structure in the large-scale flow.
They examined the conditional average $Nu$ (i.e., the average $Nu$ corresponding to a particular flow structure) and
found that the single-roll flow structure is more efficient for heat transfer than the double-roll structure.
van der Poel et al. \cite{van2011connecting,van2012flow} numerically simulated the aspect ratio dependence of $Nu$ in a two-dimensional (2D) square cell.
They conditionally averaged $Nu$ based on flow structures and found that heat transfer is more efficient with less vertically arranged vortices or less horizontally elongated vortices.
On the other hand, an interesting feature of the LSC is the spontaneous and random directional reversal, which is related to the reversal of the Earth's magnetic field \cite{petrelis2009simple} and reversal of the convective wind in the atmosphere \cite{gallet2012reversals}.
During flow reversal in the RB convection, $Nu$ first drops to its minima (corresponding to the breakup of the main roll) and then increases to its normal value (corresponding to the re-establishment of the main roll).
Xi et al. \cite{xi2016higher} experimentally observed that $Nu$ has a momentary overshoot above its average value during flow reversal.
The overshoot in $Nu$ was attributed to more coherent flow or plumes for the short period of time during reversal.
In short, a more coherent flow would produce a higher heat transfer efficiency, and thus a larger $Nu$ value.

An effective approach to extract internal flow structures from the turbulence dataset is flow mode decomposition analysis, such as Fourier mode decomposition \cite{petschel2011statistical,chandra2011dynamics} and proper orthogonal decomposition (POD) analysis \cite{lumley1967structure,berkooz1993proper}.
In these approaches, the instantaneous flow field is projected onto orthogonal basis, the instantaneous amplitude of the flow mode serves as the metrics to measure the strength of each flow mode.
The relationship between the heat transfer efficiency and each flow mode can be obtained by calculating their cross-correlation function.
A positive correlation would suggest that the flow mode produces more efficient heat transfer on average, and \emph{vice versa}.
In this work, we compare the features of four different Nusselt numbers (i.e., $Nu_{\text{vol}}$, $Nu_{\text{wall}}$,  $Nu_{\text{kinetic}}$, and $Nu_{\text{thermal}}$), particularly their abilities on revealing the connection between the heat transfer efficiency and flow structures in the RB turbulent convection.
As will become clear, proper choice of $Nu$ is critical to obtain a meaningful interpretation on how the flow structure affects the global heat transfer.
Meanwhile, we should note the advantages and disadvantages of each flow mode decomposition analysis approach.
First, in Fourier mode decomposition, we have to pre-design an appropriate Fourier basis, which may be nontrivial for complex geometry of the flow domain; in contrast, the POD does not require prior knowledge of the geometry of the flow domain.
Second, the POD modes are ranked with respect to their energy content, while the same Fourier mode can be adopted for flows with different control parameters (e.g., $Ra$ and $Pr$).

The rest of this paper is organized as follows.
In Sec. \ref{Section2}, we first present the mathematical model for the incompressible thermal flow under the Boussinesq approximation, followed by the lattice Boltzmann (LB) method to obtain velocity and temperature fields.
In Sec. \ref{Section3}, we first present general features of four different Nusselt numbers and then analyze the cross correlation between  $Nu$ and the energy of the Fourier mode, the cross correlation between $Nu$ and the amplitude of the POD mode, as well as the ensemble-averaged $Nu$ during flow reversal.
In Sec. \ref{Section4}, the main conclusions of the present work are summarized.

\section{\label{Section2}Numerical method}

\subsection{Direct numerical simulation of turbulent thermal convection}

We consider incompressible thermal flows under the Boussinesq approximation.
The temperature is treated as an active scalar, and its influence on the velocity field is realized through the buoyancy term.
The viscous heat dissipation and compression work are neglected, and all the transport coefficients are assumed to be constants.
The governing equations can be written as
\begin{subequations}
\begin{align}
& \nabla \cdot \mathbf{u}=0 \\
& \frac{\partial \mathbf{u}}{\partial t}+\mathbf{u}\cdot \nabla \mathbf{u}=-\frac{1}{\rho_{0}}\nabla p+\nu \nabla^{2}\mathbf{u}+g\beta(T-T_{0})\hat{\mathbf{z}} \\
& \frac{\partial T}{\partial t}+\mathbf{u}\cdot \nabla T=\kappa \nabla^{2} T
\end{align} \label{Eq.NS}
\end{subequations}
where $\mathbf{u}=(u,w)$ is the fluid velocity.
$p$ and $T$ are the pressure and temperature of the fluid, respectively.
$\rho_{0}$ and $T_{0}$ are the reference density and temperature, respectively.
$\hat{\mathbf{z}}$ is the unit vector in the vertical direction.
We study the flow and heat transfer in a 2D cell for two reasons. First, the computational cost for 2D simulations is much lower than that of the 3D simulations, and thus, we can adopt a fine resolution of the boundary layers to capture the extreme events at high Rayleigh numbers. Second, a particular configuration of choice in the experimental studies is the quasi-2D rectangular geometry, which enables the minimization or even elimination of the influence from the three-dimensional dynamic features of the large-scale circulation. To efficiently mimic the quasi-2D rectangular cell adopted in the experiment, we choose the 2D cell in the numerical simulation.

We adopt the lattice Boltzmann (LB) method \cite{chen1998lattice,aidun2010lattice,xu2017lattice,huang2015multiphase} as the numerical tool for DNS of turbulent thermal convection, instead of directly solving the discretized nonlinear partial differential equations.
The advantages of the LB method include easy implementation and parallelization, particularly on heterogeneous computing platforms, such as GPUs \cite{xu2017accelerated}.
The LB model to solve fluid flows and heat transfer is based on the double distribution function approach, which consists of a D2Q9 model for the Navier-Stokes equations (i.e., Eqs. \ref{Eq.NS}a and \ref{Eq.NS}b) to simulate fluid flows and a D2Q5 model for the convection-diffusion equations (i.e., Eq. \ref{Eq.NS}c) to simulate heat transfer.
In the LB method, to solve Eqs. \ref{Eq.NS}a and \ref{Eq.NS}b, the evolution equation of the density distribution function is written as
\begin{equation}
  f_{i}(\mathbf{x}+\mathbf{e}_{i}\delta_{t},t+\delta_{t})-f_{i}(\mathbf{x},t)=-(\mathbf{M}^{-1}\mathbf{S})_{ij}\left[\mathbf{m}_{j}(\mathbf{x},t)-\mathbf{m}_{j}^{(\text{eq})}(\mathbf{x},t)\right]
  +\delta_{t}F_{i}^{'} \label{Eq.MRT}
\end{equation}
where $f_{i}$ is the density distribution function.
$\mathbf{x}$ is the fluid parcel position, $t$ is the time, and $\delta_{t}$ is the time step.
$\mathbf{e}_{i}$ is the discrete velocity along the $i$th direction.
$\mathbf{M}$ is the orthogonal transformation matrix that projects the density distribution function $f_{i}$ and its equilibrium $f_{i}^{(\text{eq})}$ from the velocity space onto the moment space as $\mathbf{m}=\mathbf{M}\mathbf{f}$ and $\mathbf{m}^{(\text{eq})}=\mathbf{M}\mathbf{f}^{(\text{eq})}$.
$\mathbf{S}$ is the diagonal relaxation matrix and
$F_{i}^{'}$ is the forcing term.
The macroscopic density $\rho$ and velocity $\mathbf{u}$ are obtained from $\rho=\sum_{i=0}^{8}f_{i}$ and $\mathbf{u}=\left( \sum_{i=0}^{8}\mathbf{e}_{i}f_{i}+\mathbf{F}/2 \right)/\rho$, where $\mathbf{F}=\rho g\beta(T-T_{0})\hat{\mathbf{z}}$.
To solve Eq. \ref{Eq.NS}c, the evolution equation of temperature distribution function is written as
\begin{equation}
  g_{i}(\mathbf{x}+\mathbf{e}_{i}\delta_{t},t+\delta_{t})-g_{i}(\mathbf{x},t)=-(\mathbf{N}^{-1}\mathbf{Q})_{ij}\left[\mathbf{n}_{j}(\mathbf{x},t)-\mathbf{n}_{j}^{(\text{eq})}(\mathbf{x},t)\right]
  \label{Eq.MRT_T}
\end{equation}
where $g_{i}$ is the temperature distribution function.
$\mathbf{N}$ is the orthogonal transformation matrix that projects the temperature distribution function $g_{i}$ and its equilibrium $g_{i}^{(\text{eq})}$ from the velocity space onto the moment space as $\mathbf{n}=\mathbf{N}\mathbf{g}$ and $\mathbf{n}^{(\text{eq})}=\mathbf{N}\mathbf{g}^{(\text{eq})}$.
$\mathbf{Q}$ is the diagonal relaxation matrix.
The macroscopic temperature $T$ is obtained from $T=\sum_{i=0}^{4}g_{i}$.
More numerical details on the LB method and validation of the in-house DNS code can be found in our previous work \cite{xu2019lattice,xu2019statistics,xu2020transport}.

\subsection{\label{subsectionSettings}Simulation settings}

The top and bottom walls of the convection cell are kept at constant cold and hot temperatures, respectively, while the other two vertical walls are adiabatic.
All four walls impose no-slip velocity boundary condition.
The dimension of the cell is $H \times H$.
Simulation results are provided for the Rayleigh number of $10^{7} \le Ra \le 10^{9}$ and the fixed Prandtl number of $Pr = 5.3$.
After reaching the statistically stationary state, we take another time span of $t_{avg}$ to obtain statistically convergent results for turbulent analysis.
In Table \ref{Table1}, we list $t_{avg}$ both in  free-fall time unit  $t_{f}=\sqrt{H/(g\beta\Delta_{T})}$ and large-eddy turnover time  unit $t_{E} \approx 4 \pi/\langle |\omega_{c}(t)| \rangle_{t}$, with $\omega_{c}$ denoting the vorticity at the cell center.
Because flow reversal occurs frequently at $Ra=10^{8}$, to have enough statistics for the reversal events at $Ra=10^{8}$, we simulate as long as 480 000 $t_{f}$, which enables us to identify 694 flow reversal events.
We check whether the grid spacing $\Delta_{g}$ and time interval $\Delta_{t}$ are properly resolved by comparing with the Kolmogorov and Batchelor scales.
Here, the Kolmogorov length scale \cite{kolmogorov1941local} is estimated by the global criterion  $\eta_{K}=(\nu^{3}/\langle \varepsilon_{u} \rangle)^{1/4}=HPr^{1/2}/[Ra(Nu-1)]^{1/4}$, the Batchelor length scale  \citep{batchelor1959small,silano2010numerical} is estimated by  $\eta_{B}=\eta_{K} Pr^{-1/2}$, and the Kolmogorov time scale \cite{kolmogorov1941local} is estimated as  $\tau_{\eta}=\sqrt{\nu/\langle \varepsilon_{u} \rangle}=t_{f}\sqrt{Pr/(Nu-1)}$.
We use the volume-averaged $Nu_{\text{vol}}$ to estimate the spatial and temporal resolutions because the other three definitions of Nusselt numbers give very similar values of $Nu$, as discussed in Sec. \ref{Section31}.
From Table \ref{Table1}, we can see that grid spacings satisfy  $\max(\Delta_{g}/\eta_{K}, \Delta_{g}/\eta_{B}) \le 0.46$ and the time intervals satisfy $\Delta_{t} \le 0.00034 \tau_{\eta}$, which ensure the spatial and temporal resolutions of the DNS.

\begin{table}[h]
\caption{Spatial and temporal resolutions of the simulations.}
\begin{ruledtabular}
\begin{tabular}{cccccccc}
$Ra$	& $Pr$	& Mesh size	& $\Delta_{g}/\eta$	& $\Delta_{g}/\eta_{B}$	& $\Delta_{t}/\tau_{\eta}$	& $t_{avg}/t_{f}$ & $t_{avg}/t_{E}$ \\
\hline
$ 10^{7}$	& 5.3	& $257^{2}$	    & 0.18	& 0.41	& $3.43\times10^{-4}$	& 240 000   & 17 622  \\
$ 10^{8}$	& 5.3	& $513^{2}$	    & 0.19	& 0.44	& $2.46\times10^{-4}$	& 480 000   & 25 229  \\
$ 10^{9}$	& 5.3	& $1025^{2}$	& 0.20	& 0.46	& $1.74\times10^{-4}$	&  10 000    & 1 076 \\
\end{tabular}
\end{ruledtabular} \label{Table1}
\end{table}

\section{\label{Section3}Results and discussion}

\subsection{\label{Section31}General features of Nusselt numbers}

\begin{figure}
\centering
\includegraphics[width=12cm]{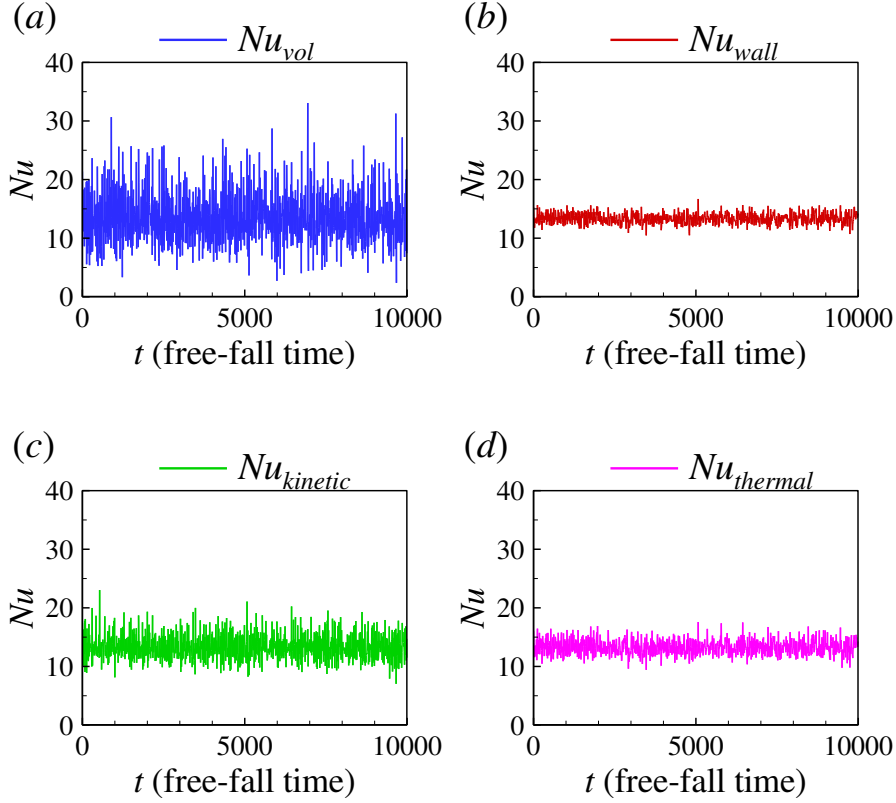}
\caption{\label{Fig1} Time series of instantaneous Nusselt numbers: (\textit{a}) the volume-averaged $Nu_{\text{vol}}$, (\textit{b}) the wall-averaged $Nu_{\text{wall}}$, (\textit{c}) the kinetic energy dissipation based $Nu_{\text{kinetic}}$, and (\textit{d}) the thermal energy dissipation based  $Nu_{\text{thermal}}$ at $Ra = 10^{7}$ and $Pr = 5.3$.}
\end{figure}

To have a general understanding of the features for the four different Nusselt numbers, we plot the time series of instantaneous Nusselt numbers at $Ra=10^{7}$ and $Pr=5.3$ for the period of 10 000 free-fall time.
From Fig. \ref{Fig1}, we can see that all the four time series have similar mean values over time, which is consistent with previous findings that different approaches to calculate the Nusselt numbers would give consistent mean values, as introduced in Sec. \ref{Section1}.
On the other hand, the volume-averaged $Nu_{\text{vol}}$ has the most significant fluctuation, while the wall-averaged $Nu_{\text{wall}}$ has the smallest fluctuation.
As for the kinetic energy dissipation based $Nu_{\text{kinetic}}$ and thermal energy dissipation based $Nu_{\text{thermal}}$, the fluctuation in the former values is larger than that in the latter one.
We can understand that as the velocity field is more intensely varied compared to the temperature field, leading to stronger temporal fluctuations in $Nu_{\text{vol}}$ and $Nu_{\text{kinetic}}$ that include the velocity field information.

\begin{figure}
\centering
\includegraphics[width=12cm]{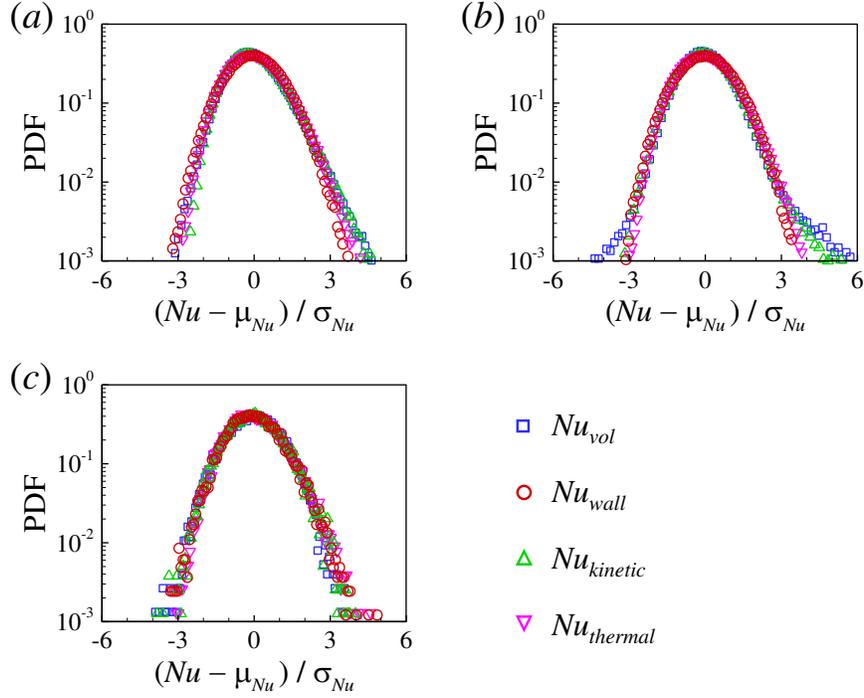}
\caption{\label{Fig2} Probability density functions (PDFs) of the normalized Nusselt numbers $(Nu-\mu_{Nu})/\sigma_{Nu}$ at (\textit{a}) $Ra = 10^{7}$, (\textit{b}) $Ra = 10^{8}$, and (\textit{c}) $Ra = 10^{9}$.}
\end{figure}

We further check the probability density functions (PDFs) of the four Nusselt numbers.
In Fig. \ref{Fig2}, we show the PDFs of the normalized Nusselt numbers $(Nu-\mu_{Nu})/\sigma_{Nu}$.
Here, $\mu_{Nu}$ and $\sigma_{Nu}$ represent the mean value and standard deviation of $Nu$.
Generally, the distributions of the normalized $Nu$ are universe, and all profiles of the PDFs collapse onto a single curve.
However, we should also note the differences in the distributions of PDFs for flows with different $Ra$.
At $Ra = 10^{7}$ and $Ra=10^{8}$, the distribution is asymmetric (right-skewed) and can be described by the gamma distribution or generalized extreme value (GEV) distribution;
in contrast, at $Ra = 10^{9}$, the distribution is symmetric and can be described by the Gaussian distribution.
A possible reason is that at $Ra = 10^{7}$ and $Ra = 10^{8}$, the LSC is unstable and reverses its direction frequently; the erratic behavoir of the LSC leads to fluctuations in $Nu$ with more extreme events \cite{aumaitre2003statistical}.
At $Ra = 10^{9}$, the LSC is much more stable, the random fluctuations of $Nu$ follow the Gaussian distribution.

We quantitatively evaluate the mean and root-mean-square (r.m.s.) values of the four instantaneous Nusselt numbers, as shown in Table \ref{Table2}.
We also provide the reference results from Zhang et al. \cite{zhang2017statisticsJFM} with the same simulation settings (denoted as $Nu_{\text{ref}}$).
The relative difference can be calculated as $|\langle Nu_{i} \rangle_{t}-Nu_{\text{ref}}|/Nu_{\text{ref}}$ with $i \in \{\text{vol}, \text{wall}, \text{kinetic}, \text{thermal} \}$, and the differences are included in brackets in the corresponding rows.
From Table \ref{Table2}, we can see that the differences in Nusselt numbers are within 1\%, indicating that our results are consistent with the previous one.
On the other hand, we can see in the present simulations, the Nusselt numbers calculated from four different approaches show good consistency with each other.
We also calculate the ratio between the r.m.s. and the mean value of the Nusselt numbers to measure their relative fluctuation as $\sigma_{Nu_{i}}/\langle Nu_{i} \rangle_{t}$, and the results are included in the brackets in the corresponding rows.
At higher $Ra$, there are more extreme events, yet occur at thinner boundary layers.
Thus, with the increase in $Ra$, the r.m.s. of $Nu$ increases, while its relative fluctuation decreases.
Overall, the relative fluctuation is significantly larger for $Nu_{\text{vol}}$ than that for $Nu_{\text{wall}}$.
\begin{table}[h]
\centering
\caption{The mean and root-mean-square (r.m.s.) values of Nusselt numbers. Data included in the brackets represent the relative difference of the Nusselt number as $|\langle Nu_{i}\rangle_{t}-Nu_{\text{ref}}|/Nu_{\text{ref}}$ or the relative fluctuation of the Nusselt number as $\sigma_{Nu_{i}}/\langle Nu_{i} \rangle_{t}$, where $i \in \{\text{vol}, \text{wall}, \text{kinetic}, \text{thermal} \}$.}
\setlength{\tabcolsep}{5mm}
\begin{ruledtabular}
\begin{tabular}{ccccccccccc}
  $Ra$& $10^{7}$ & $10^{8}$ & $10^{9}$  \\
  \hline
  $Nu_{\text{ref}}$ (Ref.~\onlinecite{zhang2017statisticsJFM}) & 13.28 & 26.21           & 51.28          \\
  $\langle Nu_{\text{vol}}\rangle_{t}$        & 13.36 (0.60\%)     & 26.36 (0.57\%)  & 51.53 (0.49\%)   \\
  $\langle Nu_{\text{wall}}\rangle_{t}$       & 13.37 (0.68\%)     & 26.38 (0.65\%)  & 51.57 (0.57\%)   \\
  $\langle Nu_{\text{kinetic}}\rangle_{t}$    & 13.31 (0.23\%)     & 26.30 (0.34\%)  & 51.46 (0.35\%)   \\
  $\langle Nu_{\text{thermal}}\rangle_{t}$    & 13.29 (0.08\%)     & 26.23 (0.08\%)  & 51.31 (0.06\%)   \\
  $\sigma_{Nu_{\text{vol}}}$             & 4.43 (33.2\%)      & 8.56 (32.5\%)      & 12.17 (23.6\%)     \\
  $\sigma_{Nu_{\text{wall}}}$            & 0.88 (6.6\%)       & 1.66 (6.3\%)       & 1.88 (3.7\%)       \\
  $\sigma_{Nu_{\text{kinetic}}}$         & 2.23 (16.8\%)      & 3.24 (12.3\%)      & 3.52 (6.8\%)       \\
  $\sigma_{Nu_{\text{thermal}}}$         & 1.30 (9.8\%)       & 2.22 (8.5\%)       & 2.68 (5.2\%)       \\
\end{tabular}
\end{ruledtabular}
\label{Table2}
\end{table}

\begin{figure}
\centering
\includegraphics[width=14cm]{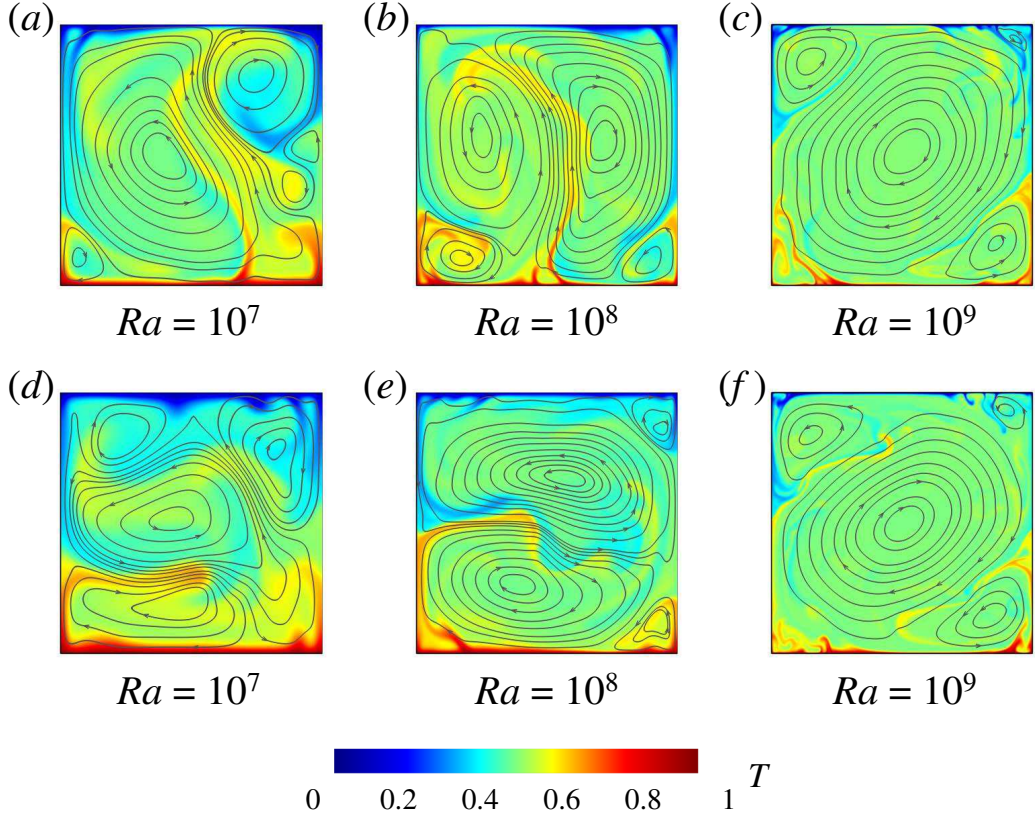}
\caption{\label{Fig3} Typical snapshots of the temperature field and streamlines when the instantaneous $Nu$ reaches  (\textit{a}-\textit{c}) 'extreme' large value  or (\textit{d}-\textit{f}) 'extreme' small value : (\textit{a} and \textit{d}) at $Ra=10^{7}$, (\textit{b} and \textit{e}) at $Ra=10^{8}$, (\textit{c} and \textit{f}) at $Ra=10^{9}$.}
\end{figure}

We then examine the flow and temperature fields when the instantaneous $Nu$ reaches 'extreme' large or small values, namely, the instant when $Nu(t) > (\langle Nu \rangle+3\sigma_{Nu})$ or $Nu(t) < (\langle Nu \rangle-3\sigma_{Nu})$.
In Fig. \ref{Fig3}, the top panel shows typical snapshots of temperature field and streamlines when the instantaneous $Nu$ reaches an 'extreme' large value, while the bottom panel shows snapshots when $Nu$ reaches an 'extreme' small value.
We can see that at $Ra=10^{7}$ and $10^{8}$, the flow structures change significantly for these two states.
Specifically, when $Nu(t) > (\langle Nu \rangle+3\sigma_{Nu})$, there exist two vertically stacked rolls, the thermal plumes rising from the hot bottom wall almost vertically hit the opposite wall.
When $Nu(t) < (\langle Nu \rangle-3\sigma_{Nu})$, there exist two horizontally stacked rolls, the plumes that are rising along the vertical wall lose their kinetic energy at half-height and then exhibit horizontal motion.
Thus, here, we provide direct evidence that  heat transfer is on average efficient with the two vertically stacked rolls, while it is on average inefficient with the two horizontally stacked rolls.
At $Ra=10^{9}$, the flow structure remains a stable big roll, which is nearly independent of the variation of $Nu$.

\subsection{\label{Section32}Cross correlation between Nusselt numbers and the Fourier mode of the flow}

\begin{figure}
\centering
\includegraphics[width=11cm]{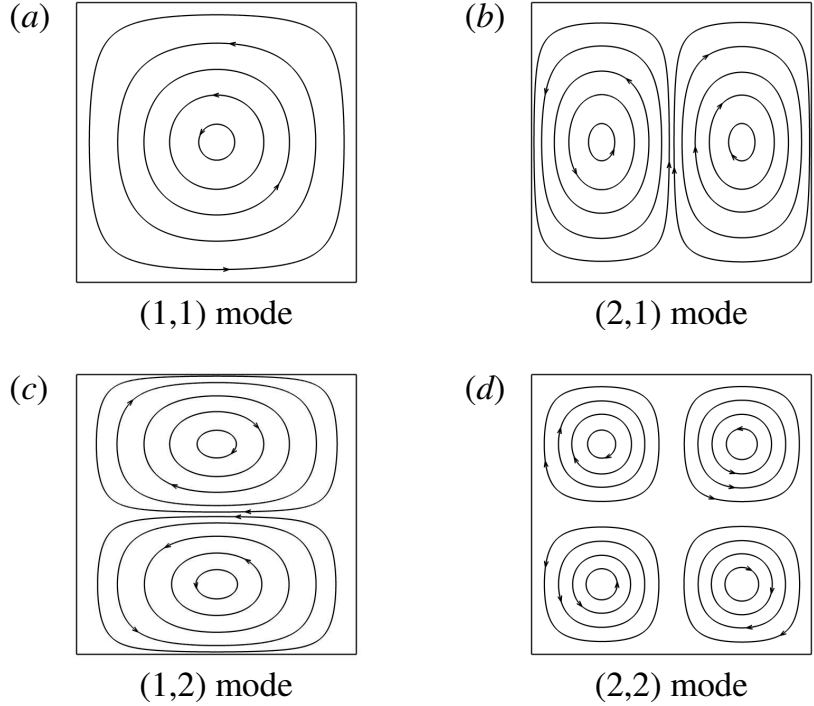}
\caption{\label{Fig4} Schematic illustration of the first four Fourier modes: (\textit{a}) the $(1, 1)$ mode, (\textit{b}) the $(2, 1)$ mode, (\textit{c}) the $(1, 2)$ mode, and (\textit{d}) the $(2, 2)$ mode.}
\end{figure}

The Fourier mode decomposition has been employed to study flow reversal mechanisms in 2D and quasi-2D square convection cells \citep{petschel2011statistical,chandra2011dynamics,chandra2013flow,verma2015flow,xi2016higher,wang2018flow,chen2019emergence}, as well as heat transfer properties in a quasi-2D cell \cite{wagner2013aspect,chong2018effect}.
Specifically, the instantaneous velocity field $(u, w)$ is projected onto the Fourier basis $(\hat{u}^{m,n},\hat{w}^{m,n})$ as
\begin{subequations}
\begin{equation}
u(x,z,t)=\sum_{m,n}A_{x}^{m,n}(t)\hat{u}^{m,n}(x,z)
\end{equation}
\begin{equation}
w(x,z,t)=\sum_{m,n}A_{z}^{m,n}(t)\hat{v}^{m,n}(x,z)
\end{equation}
\end{subequations}
Here, the Fourier basis $(\hat{u}^{m,n},\hat{w}^{m,n})$ is chosen as \citep{petschel2011statistical,chandra2011dynamics}
\begin{subequations}
\begin{equation}
\hat{u}^{m,n}(x,z)=2\sin(m\pi x)\cos(n\pi z)
\end{equation}
\begin{equation}
\hat{w}^{m,n}(x,z)=-2\cos(m\pi x)\sin(n\pi z)
\end{equation}
\end{subequations}
Although the above Fourier basis functions do not satisfy the no-slip velocity boundary condition, it was shown previously that the Fourier mode decomposition captures the convection flow profiles well \citep{petschel2011statistical,chandra2011dynamics,chandra2013flow}.
The instantaneous amplitude of the Fourier mode is then calculated as
\begin{subequations}
\begin{equation}
A_{x}^{m,n}(t)
=\langle u(x,z,t), \hat{u}^{m,n}(x,z) \rangle
=\sum_{i} \sum_{j} u(x_{i},z_{j},t)\hat{u}^{m,n}(x_{i},z_{j})
\end{equation}
\begin{equation}
A_{z}^{m,n}(t)
=\langle w(x,z,t), \hat{w}^{m,n}(x,z) \rangle
=\sum_{i} \sum_{j} w(x_{i},z_{j},t)\hat{w}^{m,n}(x_{i},z_{j})
\end{equation}
\end{subequations}
where $\langle u, \hat{u} \rangle$ and $\langle w, \hat{w} \rangle$ denote the inner product of $u$ and $\hat{u}$, $w$ and $\hat{w}$, respectively.
The energy in each Fourier mode \cite{wagner2013aspect} is evaluated as $E^{m,n}(t)=\sqrt{[A_{x}^{m,n}(t)]^{2}+[A_{z}^{m,n}(t)]^{2}}$.
Here, the  $(m,n)$ Fourier mode corresponds to a flow structure with $m$ rolls in the $x$-direction and $n$ rolls in the $z$-direction, as illustrated in Fig. \ref{Fig4}.
In the following, we will consider $m$ and $n = 1, 2, 3$, namely, the first nine Fourier modes.

\begin{figure}
\centering
\includegraphics[width=14cm]{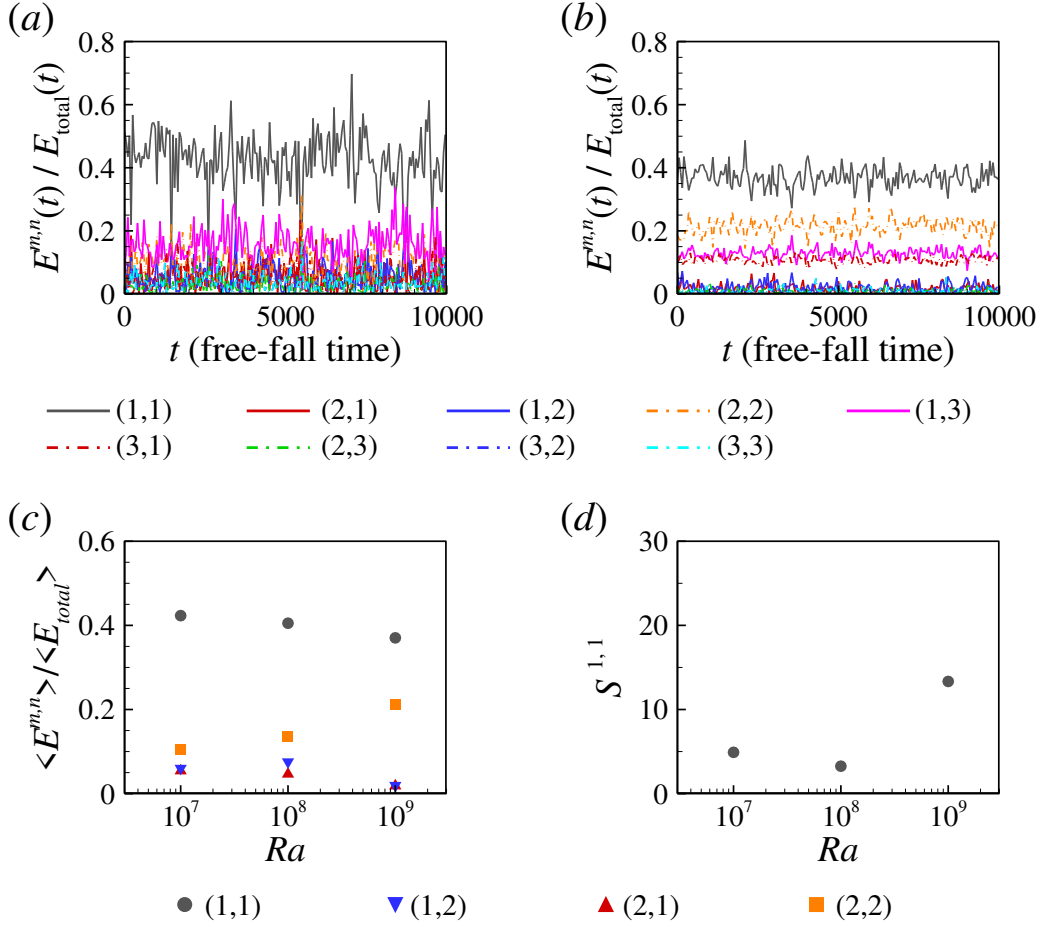}
\caption{\label{Fig5}  The time evolution of the energy in each Fourier mode for (\textit{a}) $Ra=10^{7}$ and (\textit{b}) $Ra=10^{9}$; (\textit{c}) the time-averaged energy in each Fourier mode as a function of $Ra$; (\textit{d}) the stability of the $(1,1)$ mode as a function of $Ra$.}
\end{figure}

The time evolution of energy in each Fourier mode at $Ra=10^{7}$ and $Ra=10^{9}$ for the period of 10 000 free-fall time is plotted in Figs. \ref{Fig5}(a) and \ref{Fig5}(b), respectively.
Here, we normalize the energy of the $(m,n)$ Fourier mode $E^{m,n}(t)$ by dividing the total energy $E_{total}(t)=\sum_{m,n} E^{m,n}(t)$.
We can see that for both $Ra$, the dominant Fourier mode is the $(1, 1)$ mode because it accounts for over 40\% of the total energy.
We can understand this flow mode as the primary roll in the cell center, corresponding to the large-scale circulation of the flow.
The time-averaged energy in each Fourier mode $\langle E^{m,n}(t)/E_{total}(t) \rangle_{t}$ as a function of $Ra$ is further plotted in Fig. \ref{Fig5}(c).
At $Ra=10^{9}$, the large-scale roll is of a tilde elliptical shape and it does not concentrate near the perimeter of the cell \cite{xia2003particle}, and thus the relative contribution from the $(1,1)$ mode is small and the two corner rolls account much more energy than they do in other $Ra$.
We also notice that in Figs. \ref{Fig5}(a) and \ref{Fig5}(b), the evolutions of the $(1,1)$ mode fluctuate more intense at $Ra=10^{7}$ than that at $Ra=10^{9}$.
We then calculate the stability of the $(1,1)$ Fourier mode \cite{chen2019emergence} as $S^{1,1}=\langle E^{1,1} \rangle/\sigma_{E^{1,1}}$, such that a larger value of $S^{1,1}$ indicates a more stable main roll.
We can see from Fig. \ref{Fig5}(d) that the stability of the $(1,1)$ mode is weak at $Ra=10^{7}$ and $10^{8}$, which is due to the flow reversal of the main roll.
Since the flow reversal is more frequent at $Ra=10^{8}$ compared with that at $Ra=10^{7}$, $S^{1,1}$ is smaller at the former $Ra$.
In contrast, the main roll is much more stable at $Ra=10^{9}$, which is consistent with the observations shown in Fig. \ref{Fig3}.

\begin{figure}
\centering
\includegraphics[width=14cm]{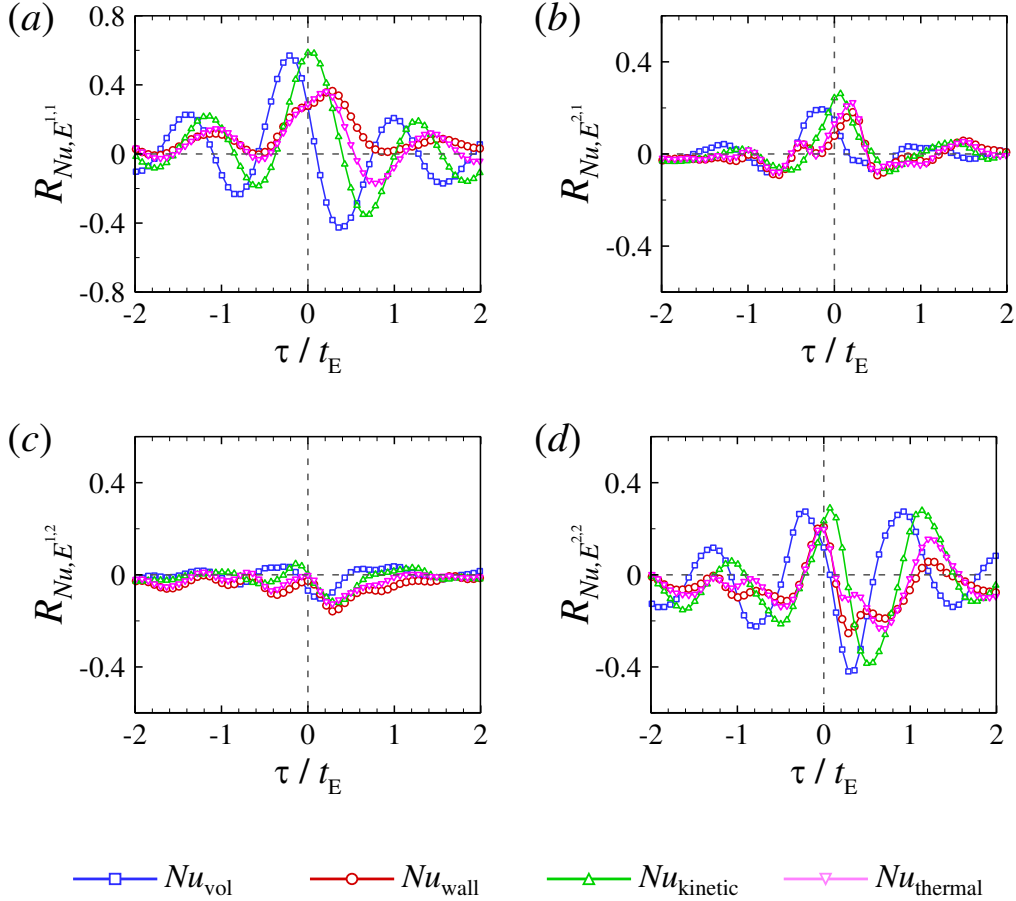}
\caption{\label{Fig6} Cross correlation between the instantaneous $Nu$ and the energy of (\textit{a}) the Fourier $(1, 1)$ mode, (\textit{b}) the Fourier $(2, 1)$ mode, (\textit{c}) the Fourier $(1, 2)$ mode, and (\textit{d}) the Fourier $(2, 2)$ mode at $Ra = 10^{7}$ and $Pr = 5.3$.}
\end{figure}

To examine the abilities of four different Nusselt numbers on revealing connections between the heat transfer efficiency and internal flow structures, we calculate the cross correlation between $Nu$ and the energy of the $(m,n)$ Fourier mode $E^{m,n}$ as $R_{Nu,E^{m,n}}(\tau)=\langle (Nu(t+\tau)-\langle Nu \rangle)(E^{m,n}(t)-\langle E^{m,n} \rangle)\rangle /(\sigma_{Nu} \sigma_{E^{m,n}})$, where $\sigma_{Nu}$ and $\sigma_{E^{m,n}}$ are the standard deviation of $Nu$ and $E^{m,n}$, respectively.
In Fig. \ref{Fig6}, we plot the cross correlation function as a function of dimensionless time delay $\tau/t_{E}$ at $Ra = 10^{7}$ and $Pr = 5.3$.
Overall, we observe periodicity in the cross correlation between instantaneous $Nu$ and energies of the $(1, 1)$ and $(2, 2)$ Fourier modes, which is due to the periodicity of these flow modes.
%%%To demonstrate how the flow structures affect the heat transfer efficiency, we only show the cross-correlation functions with a positive time delay.
From Fig. \ref{Fig6}(a), we can see that the volume-averaged $Nu_{\text{vol}}$ and the energy of the $(1,1)$ Fourier mode $E^{1,1}$ show a strong positive correlation.
Similarly, the kinetic energy dissipation based $Nu_{\text{kinetic}}$ shows a positive correlation with $E^{1,1}$ as that of $Nu_{\text{vol}}$, but only with a time lag of $\tau \approx 0.3 t_{E}$.
On the other hand, the positive correlation of wall-averaged $Nu_{\text{wall}}$ and $E^{1,1}$ is weaker compared with that of $Nu_{\text{vol}}$ and $Nu_{\text{kinetic}}$.
The thermal energy dissipation based $Nu_{\text{thermal}}$ follows a similar pattern to $Nu_{\text{wall}}$ on the correlations with $E^{1,1}$.
Previous results \cite{sun2005azimuthal,xi2008flow,van2011connecting,van2012flow} have suggested that heat transfer is more efficient with the single-roll flow structure on average, which corresponds to the $(1,1)$ Fourier mode.
Thus, the correlations of $Nu_{\text{vol}}$ and $Nu_{\text{kinetic}}$ with the $E^{1,1}$ shown here can reproduce previous findings better than $Nu_{\text{wall}}$ and $Nu_{\text{thermal}}$.
The possible reason is that both $Nu_{\text{vol}}$ and $Nu_{\text{kinetic}}$ contain velocity field information, while the flow structure obtained via Fourier flow mode analysis is also essentially based on velocity field information.
In contrast, $Nu_{\text{wall}}$ and $Nu_{\text{thermal}}$ only contain temperature field information, and they may not be good candidates to reveal the connections between the heat transfer efficiency and internal flow structures.
In Fig. \ref{Fig6}(b), we can see that all four Nusselt numbers show positive correlations with $E^{2,1}$, suggesting that the $(2,1)$ Fourier mode (corresponding to two horizontally stacked rolls) is efficient for heat transfer on average, while the only difference lies in the time lag.
In Fig. \ref{Fig6}(c),  all the Nusselt numbers show negative correlations with $E^{1,2}$, suggesting that the $(1,2)$ Fourier mode (corresponding to two vertically arranged rolls) is inefficient for heat transfer on average.
Again, the difference lies in the time lag.
Finally, for the $(2,2)$ Fourier mode, its energy is positively correlated with all the Nusselt numbers, suggesting that the quadrupolar flow is efficient for heat transfer on average.

\subsection{\label{Section33}Cross correlation between Nusselt numbers and the amplitude of POD mode}

The proper orthogonal decomposition (POD) has been employed to study flow reversal mechanisms in both 2D and 3D convection cells \citep{podvin2015large,podvin2017precursor,castillo2019cessation,soucasse2019proper}.
In the POD \cite{lumley1967structure,berkooz1993proper}, the spatiotemporal vector field $\mathbf{X}(\mathbf{r},t)$ is decomposed as a superposition of empirical orthogonal eigenfunctions $\phi_{i}(\mathbf{r})$ and their amplitudes $a_{i}(t)$ as
\begin{equation}
\mathbf{X}(\mathbf{r},t) =\sum_{i=1}^{\infty} a_{i}(t)\mathbf{\phi}_{i}(\mathbf{r})
\label{Eq.matrixX}
\end{equation}
The eigenfunctions $\mathbf{\phi}_{i}(\mathbf{r})$ are solutions of the eigenvalue problem
\begin{equation}
\int_{\Omega}
\left[ \frac{1}{N}\sum_{k=1}^{N} \mathbf{X}(\mathbf{r},t_{k})\mathbf{X}(\mathbf{r}',t_{k}) \right] \mathbf{\phi}_{i}(\mathbf{r}')d\mathbf{r}'
=\lambda_{i}\mathbf{\phi}_{i}(\mathbf{r})
\label{Eq.XX}
\end{equation}
where $\Omega$ is the spatial domain, and $N$ is the total snapshots.
If the empirical eigenfunctions are normalized, we have $\langle a_{i}(t) a_{j}(t) \rangle_{t}=\delta_{ij}\lambda_{i}$, where $\delta_{ij}$ is Kroneker symbol and $\lambda_{i}$ is the energy of the $i$th POD mode.
The eigenvalue problem described in Eq. \ref{Eq.XX} can also be written as
\begin{equation}
\mathbf{C}\Phi =   \Phi \Lambda
\label{Eq.eigenProblem}
\end{equation}
where the positive definite symmetric matrix $\mathbf{C}=(1/N)\mathbf{X}\mathbf{X}^{T}$ is the auto-correlation matrix of $\mathbf{X}$.
The columns $\phi_{i}$ of the matrix $\Phi$ are the eigenvectors of matrix $\mathbf{C}$ corresponding to the eigenvalues $\lambda_{i}$.
The matrix $\Lambda$ is a diagonal matrix containing these eigenvalues.

On the other hand, the singular value decomposition (SVD) algorithm provides a numerically stable matrix decomposition that is guaranteed to exist \cite{kutz2016dynamic,brunton2019data}.
Generally, for the dataset $\mathbf{X} \in \mathbb{R}^{n\times m}$, we have $\mathbf{X}=\mathbf{U} \Sigma \mathbf{V}^{T}$, where $\mathbf{U} \in \mathbb{R}^{n\times n}$ and $\mathbf{V} \in \mathbb{R}^{m\times m}$ are unitary matrices, and their components are denoted as $\mathbf{u}_{i}$ and $\mathbf{v}_{i}$, respectively.
$\Sigma = \text{diag}(\sigma_{1}, \sigma_{2}, \cdots, \sigma_{m})$ is a diagonal matrix containing real and non-negative entries, and these diagonal elements are singular values of matrix $\mathbf{X}$.
The SVD is closely related to the eigenvalue problem involving the correlation matrix $\mathbf{C}$.
The relation  $\mathbf{X}\mathbf{X}^{T}=\mathbf{U}\Sigma\Sigma^{T}\mathbf{U}^{T}$ indicates that the solution of the eigenvalue problem described in Eq. \ref{Eq.eigenProblem} can be solved with the SVD algorithm, where the eigenvector is $\Phi=\mathbf{U}$ and the eigenvalue is $\Lambda = (1/N)\Sigma\Sigma^{T}$.
We then have the POD mode $\phi_{i}$, its energy $\lambda_{i}$, and its the amplitude $a_{i}(t)$ as
\begin{equation}
\phi_{i}(\mathbf{r})=\mathbf{u}_{i},   \ \ \ \ \
\lambda_{i}=\frac{1}{N}\sigma_{i}^{2}, \ \ \ \ \
a_{i}(t)=\sigma_{i}\mathbf{v}_{i}
\end{equation}

\begin{figure}
\centering
\includegraphics[width=11cm]{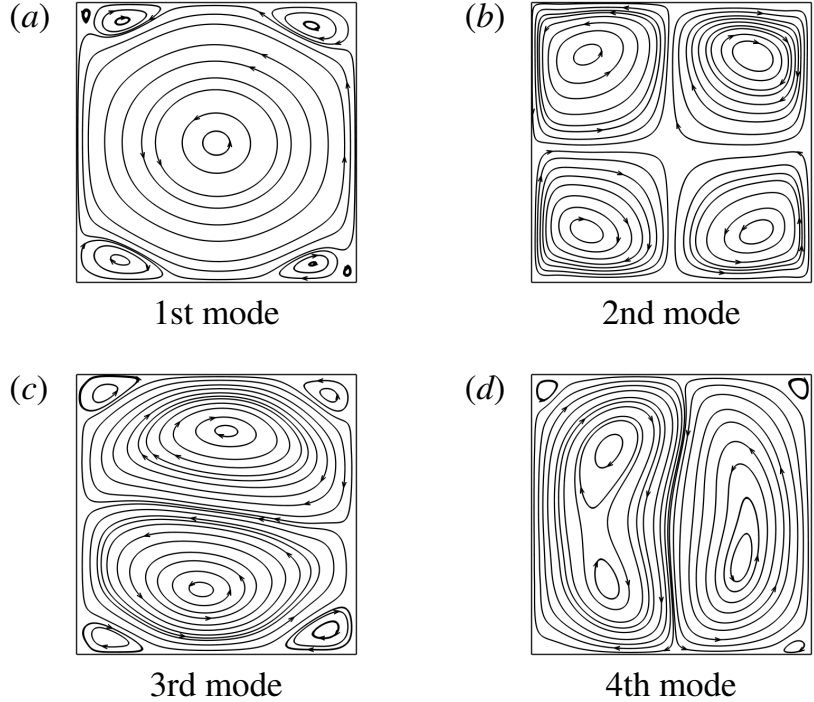}
\caption{\label{Fig7}(\textit{a}-\textit{d}) The first four proper orthogonal decomposition (POD) modes at $Ra=10^{7}$ and $Pr=5.3$ for the period of 10 000 free-fall time.}
\end{figure}

The shape of the first four POD modes is shown in Fig. \ref{Fig7}, which was obtained on a dataset of 10 000 snapshots at $Ra = 10^{7}$ and $Pr = 5.3$.
The most energetic POD mode consists of a primary roll in the cell center, which is similar to the $(1, 1)$ Fourier mode.
The second most energetic POD mode is associated with a quadrupolar flow, which corresponds to the $(2, 2)$ Fourier mode.
The third most energetic POD mode consists of two rolls stacked in the vertical direction, and it  corresponds to the $(1, 2)$ Fourier mode.
The fourth most energetic POD mode consists of two rolls stacked in the horizontal direction, and it  corresponds to the $(2, 1)$ Fourier mode.
Thus, the leading POD modes are directly related to the Fourier modes.

\begin{figure}
\centering
\includegraphics[width=14cm]{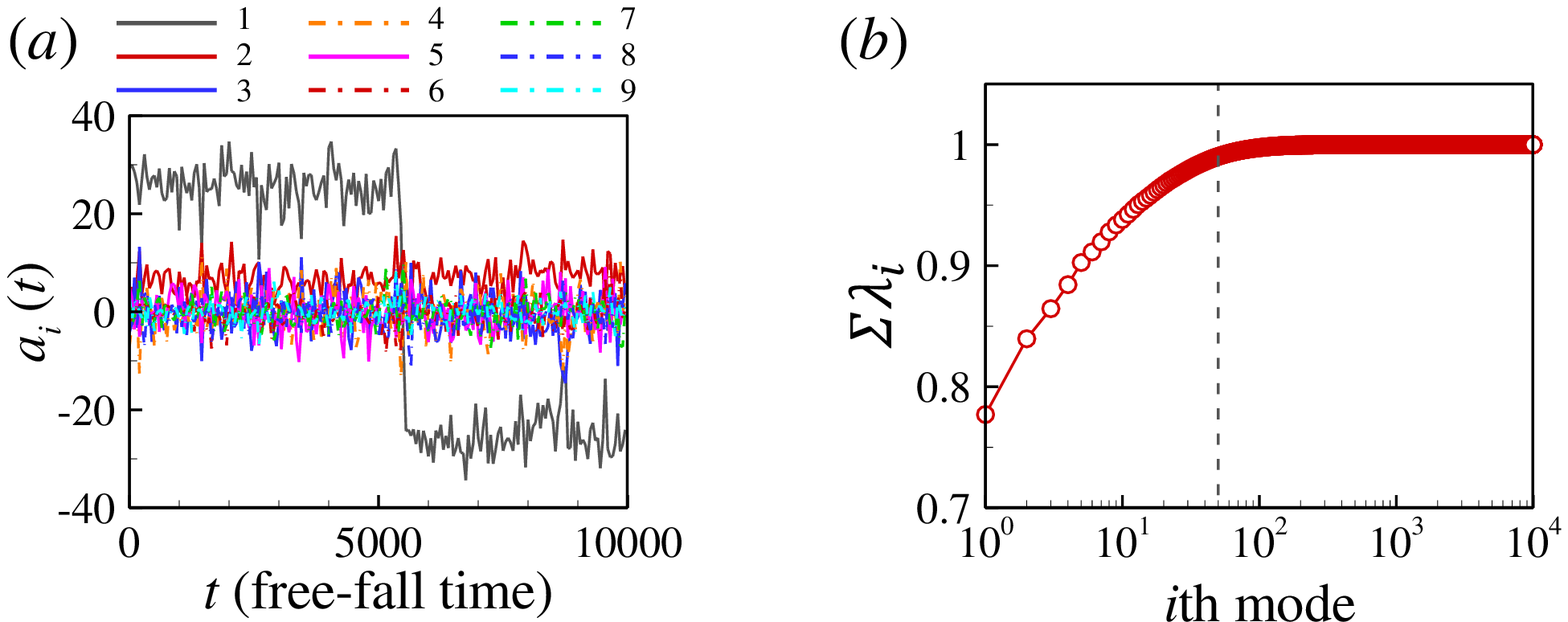}
\caption{\label{Fig8} (\textit{a}) The time evolution of the absolute value of POD mode amplitude $a_{i}(t)$ and (\textit{b}) the accumulated energy $\sum \lambda_{i}$ as a function of POD mode number $i$ at $Ra=10^{7}$ and $Pr=5.3$. The dashed gray line indicates the mode number to reach 99\% of the total energy.}
\end{figure}

The time evolution of each POD mode amplitude at $Ra=10^{7}$ and $Pr=5.3$ for the period of 10 000 free-fall time is plotted in Fig. \ref{Fig8}(a).
We can see that the amplitude of the first POD mode changes its sign at $t \approx 5500 t_{f}$, which suggests a flow reversal event because the first POD mode is related to the large-scale circulation roll.
Overall, the amplitude of the first POD mode is much larger than the rest ones, which can also be demonstrated from the accumulated energy $\sum \lambda_{i}$ as a function of POD mode number $i$.
In Fig. \ref{Fig8}(b), we can see that the first POD mode accounts for over 77\% of the total energy.
The dashed gray line in Fig. \ref{Fig8}(b) indicates that it takes 50 POD modes to reach 99\% of the total energy.

\begin{figure}
\centering
\includegraphics[width=14cm]{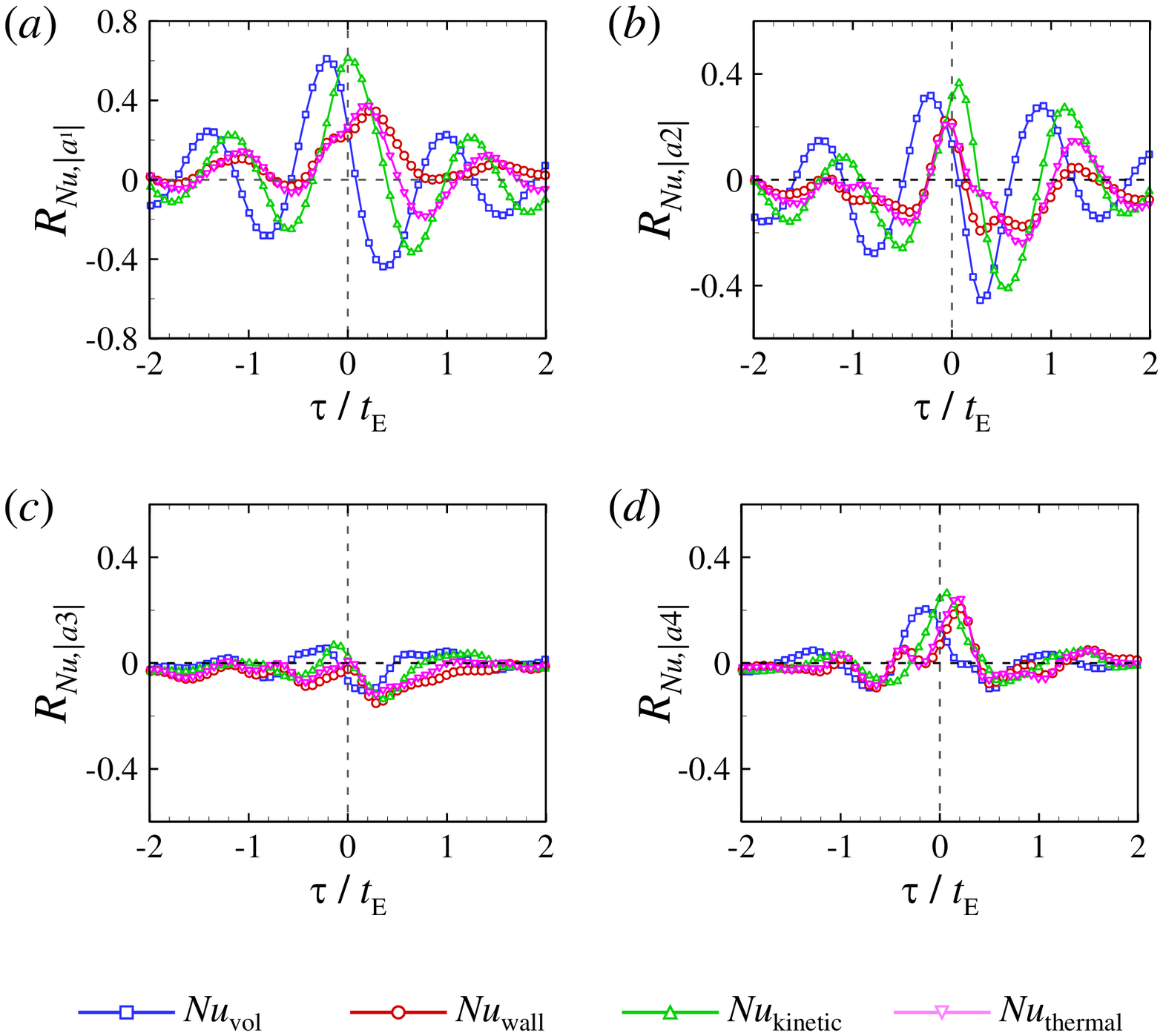}
\caption{\label{Fig9} Cross correlation between the instantaneous $Nu$ and the amplitude (absolute value) of (\textit{a}) the first POD mode, (\textit{b}) the second POD mode, (\textit{c}) the third POD mode, and (\textit{d}) the fourth POD mode at $Ra = 10^{7}$ and $Pr = 5.3$.}
\end{figure}

To further examine the abilities of four different Nusselt numbers on revealing connections between the heat transfer efficiency and internal flow structures, we then calculate the cross correlation between $Nu$ and the absolute value of POD mode amplitude $|a_{i}|$ as $R_{Nu,|a_{i}|}(\tau)=\langle (Nu(t+\tau)-\langle Nu \rangle)(|a_{i}(t)|-\langle |a_{i}| \rangle)\rangle /(\sigma_{Nu} \sigma_{|a_{i}|})$, where $\sigma_{Nu}$ and $\sigma_{|a_{i}|}$ are the standard deviation of $Nu$ and $|a_{i}|$, respectively.
Here, we adopt the absolute value of the POD mode amplitude, since the change of its sign only indicates the reversal of the flow circulation direction.
In Fig. \ref{Fig9}, we plot the cross correlation function as a function of time delay $\tau$ at $Ra = 10^{7}$ and $Pr = 5.3$.
Following the analysis on the correlation between $Nu$ and the energy of the Fourier mode $E^{m,n}$ (see Sec. \ref{Section32}), we can generally perform the same analysis on the correlation between $Nu$ and the absolute value of the POD mode amplitude $|a_{i}|$.
For the sake of clarity, we will not repeat the detailed procedure here.
The main conclusion we can draw from the results in Figs. \ref{Fig6} and \ref{Fig9} is that the POD mode dynamics exhibits almost the same behavior as that of the Fourier mode.
We can observe one-to-one correspondence in terms of the cross correlation functions between the above two different flow mode analysis approaches.
Thus, the POD analysis further justifies that using $Nu_{\text{vol}}$ and $Nu_{\text{kinetic}}$ can better reproduce the correlation between the heat transfer efficiency and flow structure obtained via flow mode analysis, and the use of these two Nusselt numbers is recommended.

\subsection{\label{Section33}Ensemble-averaged Nusselt numbers during flow reversal}
\begin{figure}
\centering
\includegraphics[width=13cm]{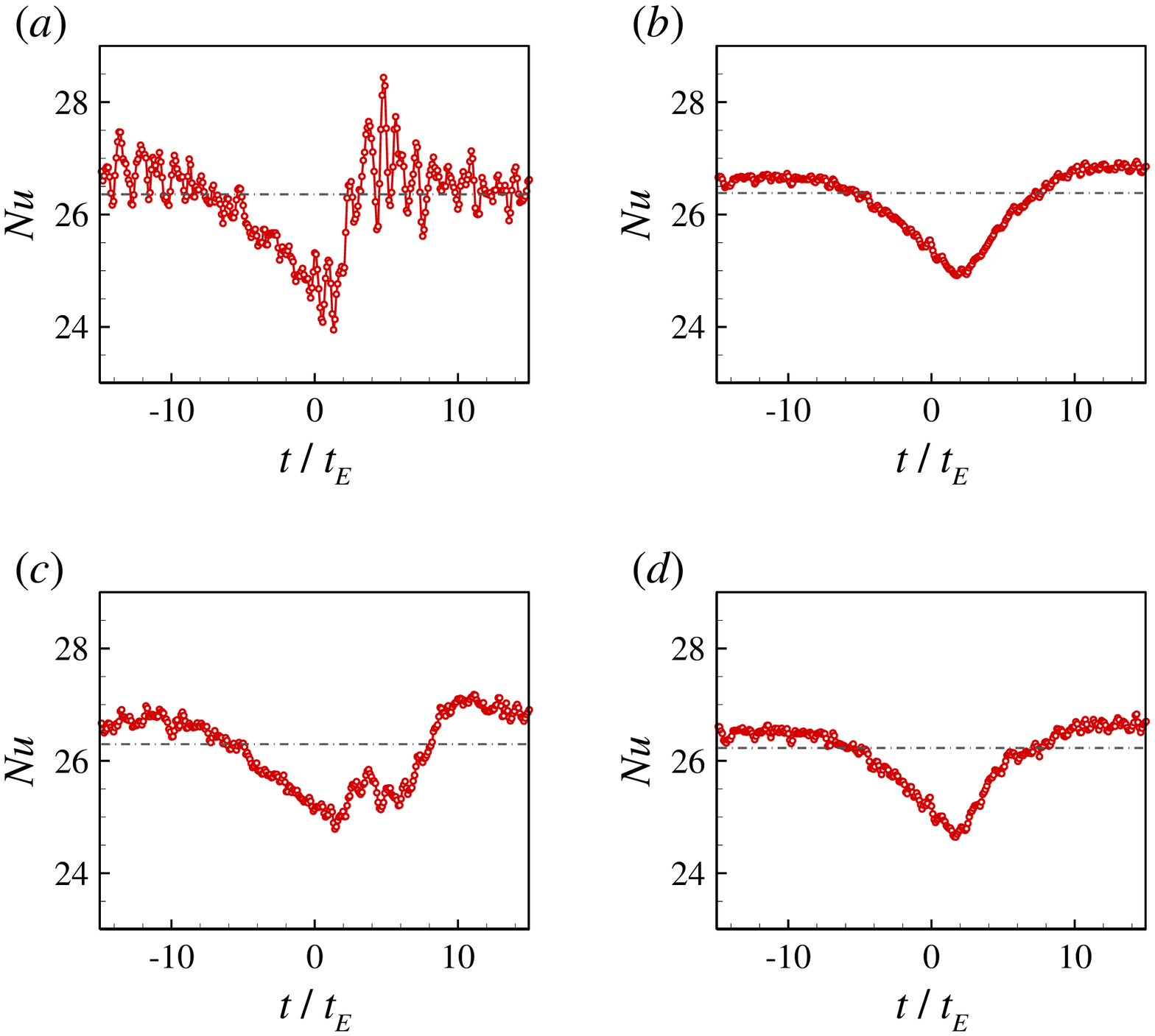}
\caption{\label{Fig10} Ensemble-averaged Nusselt numbers during flow reversal: (\textit{a}) the volume-averaged $Nu_{\text{vol}}$, (\textit{b}) the wall-averaged $Nu_{\text{wall}}$, (\textit{c}) the kinetic energy dissipation based $Nu_{\text{kinetic}}$, and (\textit{d}) the thermal energy dissipation based  $Nu_{\text{thermal}}$ at $Ra = 10^{8}$ and $Pr = 5.3$.  The dashed gray lines indicate the average value of the corresponding $Nu$.}
\end{figure}

With long-time DNS, we identified a large number of 694 reversal events at $Ra=10^{8}$ and $Pr=5.3$, which allows us to examine the behavior of different Nusselt numbers during flow reversal.
Based on the DNS data, we calculate the ensemble-averaged time trace of $Nu$ as follows:
we first locate the data point where the dimensionless angular momentum $L/L_{0}$ is crossing zero during the reversal.
Then, starting from this data point, we go forward and backward for 150 data points, respectively and extract this 300 data-point-long time segment of $Nu$ (corresponding to $30t_{E}$, which is enough to cover the mean duration time of $16t_{E}$ for the flow reversal).
After that, we average all the 694 time segments of $Nu$ onto this 300 data points, the so-obtained averaged time trace exhibits the ensemble-averaged time evolution of $Nu$ during flow reversals, as shown in Fig. \ref{Fig10}.
For the wall-averaged $Nu_{\text{wall}}$, the kinetic energy dissipation based $Nu_{\text{kinetic}}$, and the thermal energy dissipation based $Nu_{\text{thermal}}$ (see Figs. \ref{Fig10}b-\ref{Fig10}d), we can only observe that the $Nu$ numbers decrease before the reversal, drop to their minima at $t \approx 0$, and then increase to their normal value after the reversal.
In contrast, we can observe a momentary overshoot in the volume-averaged $Nu_{\text{vol}}$ above its average value (see Fig. \ref{Fig10}a) during flow reversal.
We note among the four different Nusselt numbers, only $Nu_{\text{vol}}$ simultaneously includes the velocity and temperature field information, while the other three $Nu$ reflect either the velocity field or the temperature field information, which may be the reason that only $Nu_{\text{vol}}$ can reproduce the 'overshoot' phenomena observed in a previous experimental study \cite{xi2016higher}.

\section{\label{Section4}Conclusions}
In this work, we have performed high-resolution and long-time direct numerical simulations of turbulent Rayleigh-B\'enard convection to investigate the correlation of the internal flow structure and heat transfer efficiency.
Specifically, we examined the abilities of four different Nusselt numbers (i.e., the volume-averaged $Nu_{\text{vol}}$, the wall-averaged $Nu_{\text{wall}}$, the kinetic energy dissipation based $Nu_{\text{kinetic}}$, and the thermal energy dissipation based $Nu_{\text{thermal}}$) on revealing this connection.
The main findings are summarized as follows:
\begin{enumerate}
\item
All the four different Nusselt numbers exhibit consistent time-averaged mean values, and their PDFs collapse onto a single curve.
$Nu_{\text{vol}}$ shows the largest fluctuation, while $Nu_{\text{wall}}$ shows the smallest fluctuation.

\item
The Fourier mode decomposition and the POD analysis show that in the 2D square RB cell, the single-roll flow structure, the horizontally stacked roll flow structure, and  the quadrupolar flow structure are more efficient for heat transfer on average.
In contrast, the vertically stacked roll flow structure is inefficient for heat transfer on average.

\item
The cross correlation functions between instantaneous $Nu$ and flow mode amplitude indicate that $Nu_{\text{vol}}$ and $Nu_{\text{kinetic}}$ can better reproduce the correlation between the flow structure and heat transfer efficiency than that of $Nu_{\text{wall}}$ and $Nu_{\text{thermal}}$.
To analyze the correlation between $Nu$ and the flow structures obtained via flow mode analysis, we recommend using the former two Nusselt numbers.

\item
During flow reversal, a previous experimental study reported that $Nu$ has a momentary overshoot above its average value due to more coherent flow or plumes.
Among the four Nusselt numbers, only the ensemble-averaged time trace of $Nu_{\text{vol}}$  can reproduce the overshoot phenomena.
\end{enumerate}

%%\section*{\label{SectionSM}supplementary material}
%%
%%See the supplementary material for data fitting probability density functions (PDFs) of the Nusselt numbers at $10^{7} \le Ra \le 10^{9}$ and $Pr=5.3$.

\begin{acknowledgments}
This work was supported by the National Natural Science Foundation of China (NSFC) through Grant Nos. 11902268 and 11772259,
the National Key Project GJXM92579,
the Fundamental Research Funds for the Central Universities of China (Nos. D5000200570 and 3102019PJ002),
and the 111 project of China (No. B17037).
%%%The simulations were carried out at LvLiang Cloud Computing Center of China, and the calculations were performed on TianHe-2.
\end{acknowledgments}

\section*{Data Availability Statement}
The data that support the findings of this study are available from the corresponding authors upon reasonable request.

\nocite{*}
\bibliography{myBib}% Produces the bibliography via BibTeX.

\begin{thebibliography}{56}%
\makeatletter
\providecommand \@ifxundefined [1]{%
 \@ifx{#1\undefined}
}%
\providecommand \@ifnum [1]{%
 \ifnum #1\expandafter \@firstoftwo
 \else \expandafter \@secondoftwo
 \fi
}%
\providecommand \@ifx [1]{%
 \ifx #1\expandafter \@firstoftwo
 \else \expandafter \@secondoftwo
 \fi
}%
\providecommand \natexlab [1]{#1}%
\providecommand \enquote  [1]{``#1''}%
\providecommand \bibnamefont  [1]{#1}%
\providecommand \bibfnamefont [1]{#1}%
\providecommand \citenamefont [1]{#1}%
\providecommand \href@noop [0]{\@secondoftwo}%
\providecommand \href [0]{\begingroup \@sanitize@url \@href}%
\providecommand \@href[1]{\@@startlink{#1}\@@href}%
\providecommand \@@href[1]{\endgroup#1\@@endlink}%
\providecommand \@sanitize@url [0]{\catcode `\\12\catcode `\$12\catcode
  `\&12\catcode `\#12\catcode `\^12\catcode `\_12\catcode `\%12\relax}%
\providecommand \@@startlink[1]{}%
\providecommand \@@endlink[0]{}%
\providecommand \url  [0]{\begingroup\@sanitize@url \@url }%
\providecommand \@url [1]{\endgroup\@href {#1}{\urlprefix }}%
\providecommand \urlprefix  [0]{URL }%
\providecommand \Eprint [0]{\href }%
\providecommand \doibase [0]{http://dx.doi.org/}%
\providecommand \selectlanguage [0]{\@gobble}%
\providecommand \bibinfo  [0]{\@secondoftwo}%
\providecommand \bibfield  [0]{\@secondoftwo}%
\providecommand \translation [1]{[#1]}%
\providecommand \BibitemOpen [0]{}%
\providecommand \bibitemStop [0]{}%
\providecommand \bibitemNoStop [0]{.\EOS\space}%
\providecommand \EOS [0]{\spacefactor3000\relax}%
\providecommand \BibitemShut  [1]{\csname bibitem#1\endcsname}%
\let\auto@bib@innerbib\@empty
%</preamble>
\bibitem [{\citenamefont {Ahlers}, \citenamefont {Grossmann},\ and\
  \citenamefont {Lohse}(2009)}]{ahlers2009heat}%
  \BibitemOpen
  \bibfield  {author} {\bibinfo {author} {\bibfnamefont {G.}~\bibnamefont
  {Ahlers}}, \bibinfo {author} {\bibfnamefont {S.}~\bibnamefont {Grossmann}}, \
  and\ \bibinfo {author} {\bibfnamefont {D.}~\bibnamefont {Lohse}},\ }\bibfield
   {title} {\enquote {\bibinfo {title} {Heat transfer and large scale dynamics
  in turbulent {R}ayleigh-{B}{\'e}nard convection},}\ }\href {\doibase
  10.1103/RevModPhys.81.503} {\bibfield  {journal} {\bibinfo  {journal}
  {Reviews of Modern Physics}\ }\textbf {\bibinfo {volume} {81}},\ \bibinfo
  {pages} {503} (\bibinfo {year} {2009})}\BibitemShut {NoStop}%
\bibitem [{\citenamefont {Lohse}\ and\ \citenamefont
  {Xia}(2010)}]{lohse2010small}%
  \BibitemOpen
  \bibfield  {author} {\bibinfo {author} {\bibfnamefont {D.}~\bibnamefont
  {Lohse}}\ and\ \bibinfo {author} {\bibfnamefont {K.-Q.}\ \bibnamefont
  {Xia}},\ }\bibfield  {title} {\enquote {\bibinfo {title} {Small-scale
  properties of turbulent {R}ayleigh-{B}{\'e}nard convection},}\ }\href
  {\doibase 10.1146/annurev.fluid.010908.165152} {\bibfield  {journal}
  {\bibinfo  {journal} {Annual Review of Fluid Mechanics}\ }\textbf {\bibinfo
  {volume} {42}},\ \bibinfo {pages} {335--364} (\bibinfo {year}
  {2010})}\BibitemShut {NoStop}%
\bibitem [{\citenamefont {Chill{\`a}}\ and\ \citenamefont
  {Schumacher}(2012)}]{chilla2012new}%
  \BibitemOpen
  \bibfield  {author} {\bibinfo {author} {\bibfnamefont {F.}~\bibnamefont
  {Chill{\`a}}}\ and\ \bibinfo {author} {\bibfnamefont {J.}~\bibnamefont
  {Schumacher}},\ }\bibfield  {title} {\enquote {\bibinfo {title} {New
  perspectives in turbulent {R}ayleigh-{B}{\'e}nard convection},}\ }\href
  {\doibase 10.1140/epje/i2012-12058-1} {\bibfield  {journal} {\bibinfo
  {journal} {The European Physical Journal E}\ }\textbf {\bibinfo {volume}
  {35}},\ \bibinfo {pages} {58} (\bibinfo {year} {2012})}\BibitemShut {NoStop}%
\bibitem [{\citenamefont {Xia}(2013)}]{xia2013current}%
  \BibitemOpen
  \bibfield  {author} {\bibinfo {author} {\bibfnamefont {K.-Q.}\ \bibnamefont
  {Xia}},\ }\bibfield  {title} {\enquote {\bibinfo {title} {Current trends and
  future directions in turbulent thermal convection},}\ }\href {\doibase
  10.1063/2.1305201} {\bibfield  {journal} {\bibinfo  {journal} {Theoretical
  and Applied Mechanics Letters}\ }\textbf {\bibinfo {volume} {3}},\ \bibinfo
  {pages} {052001} (\bibinfo {year} {2013})}\BibitemShut {NoStop}%
\bibitem [{\citenamefont {Mazzino}(2017)}]{mazzino2017two}%
  \BibitemOpen
  \bibfield  {author} {\bibinfo {author} {\bibfnamefont {A.}~\bibnamefont
  {Mazzino}},\ }\bibfield  {title} {\enquote {\bibinfo {title} {Two-dimensional
  turbulent convection},}\ }\href {\doibase 10.1063/1.4990083} {\bibfield
  {journal} {\bibinfo  {journal} {Physics of Fluids}\ }\textbf {\bibinfo
  {volume} {29}},\ \bibinfo {pages} {111102} (\bibinfo {year}
  {2017})}\BibitemShut {NoStop}%
\bibitem [{\citenamefont {Wang}, \citenamefont {Zhou},\ and\ \citenamefont
  {Sun}(2020)}]{wang2020vibration}%
  \BibitemOpen
  \bibfield  {author} {\bibinfo {author} {\bibfnamefont {B.-F.}\ \bibnamefont
  {Wang}}, \bibinfo {author} {\bibfnamefont {Q.}~\bibnamefont {Zhou}}, \ and\
  \bibinfo {author} {\bibfnamefont {C.}~\bibnamefont {Sun}},\ }\bibfield
  {title} {\enquote {\bibinfo {title} {Vibration-induced boundary-layer
  destabilization achieves massive heat-transport enhancement},}\ }\href
  {\doibase 10.1126/sciadv.aaz8239} {\bibfield  {journal} {\bibinfo  {journal}
  {Science Advances}\ }\textbf {\bibinfo {volume} {6}},\ \bibinfo {pages}
  {eaaz8239} (\bibinfo {year} {2020})}\BibitemShut {NoStop}%
\bibitem [{\citenamefont {Heslot}, \citenamefont {Castaing},\ and\
  \citenamefont {Libchaber}(1987)}]{heslot1987transitions}%
  \BibitemOpen
  \bibfield  {author} {\bibinfo {author} {\bibfnamefont {F.}~\bibnamefont
  {Heslot}}, \bibinfo {author} {\bibfnamefont {B.}~\bibnamefont {Castaing}}, \
  and\ \bibinfo {author} {\bibfnamefont {A.}~\bibnamefont {Libchaber}},\
  }\bibfield  {title} {\enquote {\bibinfo {title} {Transitions to turbulence in
  helium gas},}\ }\href {\doibase 10.1103/PhysRevA.36.5870} {\bibfield
  {journal} {\bibinfo  {journal} {Physical Review A}\ }\textbf {\bibinfo
  {volume} {36}},\ \bibinfo {pages} {5870} (\bibinfo {year}
  {1987})}\BibitemShut {NoStop}%
\bibitem [{\citenamefont {Castaing}\ \emph {et~al.}(1989)\citenamefont
  {Castaing}, \citenamefont {Gunaratne}, \citenamefont {Heslot}, \citenamefont
  {Kadanoff}, \citenamefont {Libchaber}, \citenamefont {Thomae}, \citenamefont
  {Wu}, \citenamefont {Zaleski},\ and\ \citenamefont
  {Zanetti}}]{castaing1989scaling}%
  \BibitemOpen
  \bibfield  {author} {\bibinfo {author} {\bibfnamefont {B.}~\bibnamefont
  {Castaing}}, \bibinfo {author} {\bibfnamefont {G.}~\bibnamefont {Gunaratne}},
  \bibinfo {author} {\bibfnamefont {F.}~\bibnamefont {Heslot}}, \bibinfo
  {author} {\bibfnamefont {L.}~\bibnamefont {Kadanoff}}, \bibinfo {author}
  {\bibfnamefont {A.}~\bibnamefont {Libchaber}}, \bibinfo {author}
  {\bibfnamefont {S.}~\bibnamefont {Thomae}}, \bibinfo {author} {\bibfnamefont
  {X.-Z.}\ \bibnamefont {Wu}}, \bibinfo {author} {\bibfnamefont
  {S.}~\bibnamefont {Zaleski}}, \ and\ \bibinfo {author} {\bibfnamefont
  {G.}~\bibnamefont {Zanetti}},\ }\bibfield  {title} {\enquote {\bibinfo
  {title} {Scaling of hard thermal turbulence in {R}ayleigh-{B}{\'e}nard
  convection},}\ }\href {\doibase 10.1017/S0022112089001643} {\bibfield
  {journal} {\bibinfo  {journal} {Journal of Fluid Mechanics}\ }\textbf
  {\bibinfo {volume} {204}},\ \bibinfo {pages} {1--30} (\bibinfo {year}
  {1989})}\BibitemShut {NoStop}%
\bibitem [{\citenamefont {Xia}\ and\ \citenamefont
  {Lui}(1997)}]{xia1997turbulent}%
  \BibitemOpen
  \bibfield  {author} {\bibinfo {author} {\bibfnamefont {K.-Q.}\ \bibnamefont
  {Xia}}\ and\ \bibinfo {author} {\bibfnamefont {S.-L.}\ \bibnamefont {Lui}},\
  }\bibfield  {title} {\enquote {\bibinfo {title} {Turbulent thermal convection
  with an obstructed sidewall},}\ }\href@noop {} {\bibfield  {journal}
  {\bibinfo  {journal} {Physical Review Letters}\ }\textbf {\bibinfo {volume}
  {79}},\ \bibinfo {pages} {5006} (\bibinfo {year} {1997})}\BibitemShut
  {NoStop}%
\bibitem [{\citenamefont {Shang}, \citenamefont {Tong},\ and\ \citenamefont
  {Xia}(2008)}]{shang2008scaling}%
  \BibitemOpen
  \bibfield  {author} {\bibinfo {author} {\bibfnamefont {X.-D.}\ \bibnamefont
  {Shang}}, \bibinfo {author} {\bibfnamefont {P.}~\bibnamefont {Tong}}, \ and\
  \bibinfo {author} {\bibfnamefont {K.-Q.}\ \bibnamefont {Xia}},\ }\bibfield
  {title} {\enquote {\bibinfo {title} {Scaling of the local convective heat
  flux in turbulent {R}ayleigh-{B}{\'e}nard convection},}\ }\href@noop {}
  {\bibfield  {journal} {\bibinfo  {journal} {Physical Review Letters}\
  }\textbf {\bibinfo {volume} {100}},\ \bibinfo {pages} {244503} (\bibinfo
  {year} {2008})}\BibitemShut {NoStop}%
\bibitem [{\citenamefont {Yang}\ \emph {et~al.}(2020)\citenamefont {Yang},
  \citenamefont {Zhu}, \citenamefont {Wang}, \citenamefont {Liu},\ and\
  \citenamefont {Zhou}}]{yang2020experimental}%
  \BibitemOpen
  \bibfield  {author} {\bibinfo {author} {\bibfnamefont {Y.-H.}\ \bibnamefont
  {Yang}}, \bibinfo {author} {\bibfnamefont {X.}~\bibnamefont {Zhu}}, \bibinfo
  {author} {\bibfnamefont {B.-F.}\ \bibnamefont {Wang}}, \bibinfo {author}
  {\bibfnamefont {Y.-L.}\ \bibnamefont {Liu}}, \ and\ \bibinfo {author}
  {\bibfnamefont {Q.}~\bibnamefont {Zhou}},\ }\bibfield  {title} {\enquote
  {\bibinfo {title} {Experimental investigation of turbulent
  {R}ayleigh-{B}{\'e}nard convection of water in a cylindrical cell: {T}he
  {P}randtl number effects for {P}r > 1},}\ }\href@noop {} {\bibfield
  {journal} {\bibinfo  {journal} {Physics of Fluids}\ }\textbf {\bibinfo
  {volume} {32}},\ \bibinfo {pages} {015101} (\bibinfo {year}
  {2020})}\BibitemShut {NoStop}%
\bibitem [{\citenamefont {Kerr}(1996)}]{kerr1996rayleigh}%
  \BibitemOpen
  \bibfield  {author} {\bibinfo {author} {\bibfnamefont {R.~M.}\ \bibnamefont
  {Kerr}},\ }\bibfield  {title} {\enquote {\bibinfo {title} {Rayleigh number
  scaling in numerical convection},}\ }\href {\doibase
  10.1017/S0022112096001760} {\bibfield  {journal} {\bibinfo  {journal}
  {Journal of Fluid Mechanics}\ }\textbf {\bibinfo {volume} {310}},\ \bibinfo
  {pages} {139--179} (\bibinfo {year} {1996})}\BibitemShut {NoStop}%
\bibitem [{\citenamefont {Verzicco}\ and\ \citenamefont
  {Camussi}(2003)}]{verzicco2003numerical}%
  \BibitemOpen
  \bibfield  {author} {\bibinfo {author} {\bibfnamefont {R.}~\bibnamefont
  {Verzicco}}\ and\ \bibinfo {author} {\bibfnamefont {R.}~\bibnamefont
  {Camussi}},\ }\bibfield  {title} {\enquote {\bibinfo {title} {Numerical
  experiments on strongly turbulent thermal convection in a slender cylindrical
  cell},}\ }\href {\doibase 10.1017/S0022112002003063} {\bibfield  {journal}
  {\bibinfo  {journal} {Journal of Fluid Mechanics}\ }\textbf {\bibinfo
  {volume} {477}},\ \bibinfo {pages} {19--49} (\bibinfo {year}
  {2003})}\BibitemShut {NoStop}%
\bibitem [{\citenamefont {Shraiman}\ and\ \citenamefont
  {Siggia}(1990)}]{shraiman1990heat}%
  \BibitemOpen
  \bibfield  {author} {\bibinfo {author} {\bibfnamefont {B.~I.}\ \bibnamefont
  {Shraiman}}\ and\ \bibinfo {author} {\bibfnamefont {E.~D.}\ \bibnamefont
  {Siggia}},\ }\bibfield  {title} {\enquote {\bibinfo {title} {Heat transport
  in high-{R}ayleigh-number convection},}\ }\href {\doibase
  10.1103/PhysRevA.42.3650} {\bibfield  {journal} {\bibinfo  {journal}
  {Physical Review A}\ }\textbf {\bibinfo {volume} {42}},\ \bibinfo {pages}
  {3650} (\bibinfo {year} {1990})}\BibitemShut {NoStop}%
\bibitem [{\citenamefont {Siggia}(1994)}]{siggia1994high}%
  \BibitemOpen
  \bibfield  {author} {\bibinfo {author} {\bibfnamefont {E.~D.}\ \bibnamefont
  {Siggia}},\ }\bibfield  {title} {\enquote {\bibinfo {title} {High {R}ayleigh
  number convection},}\ }\href {\doibase 10.1146/annurev.fl.26.010194.001033}
  {\bibfield  {journal} {\bibinfo  {journal} {Annual Review of Fluid
  Mechanics}\ }\textbf {\bibinfo {volume} {26}},\ \bibinfo {pages} {137--168}
  (\bibinfo {year} {1994})}\BibitemShut {NoStop}%
\bibitem [{\citenamefont {Grossmann}\ and\ \citenamefont
  {Lohse}(2000)}]{grossmann2000scaling}%
  \BibitemOpen
  \bibfield  {author} {\bibinfo {author} {\bibfnamefont {S.}~\bibnamefont
  {Grossmann}}\ and\ \bibinfo {author} {\bibfnamefont {D.}~\bibnamefont
  {Lohse}},\ }\bibfield  {title} {\enquote {\bibinfo {title} {Scaling in
  thermal convection: a unifying theory},}\ }\href {\doibase
  10.1017/S0022112099007545} {\bibfield  {journal} {\bibinfo  {journal}
  {Journal of Fluid Mechanics}\ }\textbf {\bibinfo {volume} {407}},\ \bibinfo
  {pages} {27--56} (\bibinfo {year} {2000})}\BibitemShut {NoStop}%
\bibitem [{\citenamefont {Grossmann}\ and\ \citenamefont
  {Lohse}(2004)}]{grossmann2004fluctuations}%
  \BibitemOpen
  \bibfield  {author} {\bibinfo {author} {\bibfnamefont {S.}~\bibnamefont
  {Grossmann}}\ and\ \bibinfo {author} {\bibfnamefont {D.}~\bibnamefont
  {Lohse}},\ }\bibfield  {title} {\enquote {\bibinfo {title} {Fluctuations in
  turbulent {R}ayleigh--{B}{\'e}nard convection: the role of plumes},}\ }\href
  {\doibase 10.1063/1.1807751} {\bibfield  {journal} {\bibinfo  {journal}
  {Physics of Fluids}\ }\textbf {\bibinfo {volume} {16}},\ \bibinfo {pages}
  {4462--4472} (\bibinfo {year} {2004})}\BibitemShut {NoStop}%
\bibitem [{\citenamefont {Kooij}\ \emph {et~al.}(2018)\citenamefont {Kooij},
  \citenamefont {Botchev}, \citenamefont {Frederix}, \citenamefont {Geurts},
  \citenamefont {Horn}, \citenamefont {Lohse}, \citenamefont {van~der Poel},
  \citenamefont {Shishkina}, \citenamefont {Stevens},\ and\ \citenamefont
  {Verzicco}}]{kooij2018comparison}%
  \BibitemOpen
  \bibfield  {author} {\bibinfo {author} {\bibfnamefont {G.~L.}\ \bibnamefont
  {Kooij}}, \bibinfo {author} {\bibfnamefont {M.~A.}\ \bibnamefont {Botchev}},
  \bibinfo {author} {\bibfnamefont {E.~M.}\ \bibnamefont {Frederix}}, \bibinfo
  {author} {\bibfnamefont {B.~J.}\ \bibnamefont {Geurts}}, \bibinfo {author}
  {\bibfnamefont {S.}~\bibnamefont {Horn}}, \bibinfo {author} {\bibfnamefont
  {D.}~\bibnamefont {Lohse}}, \bibinfo {author} {\bibfnamefont {E.~P.}\
  \bibnamefont {van~der Poel}}, \bibinfo {author} {\bibfnamefont
  {O.}~\bibnamefont {Shishkina}}, \bibinfo {author} {\bibfnamefont {R.~J.}\
  \bibnamefont {Stevens}}, \ and\ \bibinfo {author} {\bibfnamefont
  {R.}~\bibnamefont {Verzicco}},\ }\bibfield  {title} {\enquote {\bibinfo
  {title} {Comparison of computational codes for direct numerical simulations
  of turbulent {R}ayleigh--{B}{\'e}nard convection},}\ }\href {\doibase
  10.1016/j.compfluid.2018.01.010} {\bibfield  {journal} {\bibinfo  {journal}
  {Computers \& Fluids}\ }\textbf {\bibinfo {volume} {166}},\ \bibinfo {pages}
  {1--8} (\bibinfo {year} {2018})}\BibitemShut {NoStop}%
\bibitem [{\citenamefont {Sun}, \citenamefont {Xi},\ and\ \citenamefont
  {Xia}(2005)}]{sun2005azimuthal}%
  \BibitemOpen
  \bibfield  {author} {\bibinfo {author} {\bibfnamefont {C.}~\bibnamefont
  {Sun}}, \bibinfo {author} {\bibfnamefont {H.-D.}\ \bibnamefont {Xi}}, \ and\
  \bibinfo {author} {\bibfnamefont {K.-Q.}\ \bibnamefont {Xia}},\ }\bibfield
  {title} {\enquote {\bibinfo {title} {Azimuthal symmetry, flow dynamics, and
  heat transport in turbulent thermal convection in a cylinder with an aspect
  ratio of 0.5},}\ }\href {\doibase 10.1103/PhysRevLett.95.074502} {\bibfield
  {journal} {\bibinfo  {journal} {Physical Review Letters}\ }\textbf {\bibinfo
  {volume} {95}},\ \bibinfo {pages} {074502} (\bibinfo {year}
  {2005})}\BibitemShut {NoStop}%
\bibitem [{\citenamefont {Xi}\ and\ \citenamefont {Xia}(2008)}]{xi2008flow}%
  \BibitemOpen
  \bibfield  {author} {\bibinfo {author} {\bibfnamefont {H.-D.}\ \bibnamefont
  {Xi}}\ and\ \bibinfo {author} {\bibfnamefont {K.-Q.}\ \bibnamefont {Xia}},\
  }\bibfield  {title} {\enquote {\bibinfo {title} {Flow mode transitions in
  turbulent thermal convection},}\ }\href {\doibase 10.1063/1.2920444}
  {\bibfield  {journal} {\bibinfo  {journal} {Physics of Fluids}\ }\textbf
  {\bibinfo {volume} {20}},\ \bibinfo {pages} {055104} (\bibinfo {year}
  {2008})}\BibitemShut {NoStop}%
\bibitem [{\citenamefont {Weiss}\ and\ \citenamefont
  {Ahlers}(2011)}]{weiss2011turbulent}%
  \BibitemOpen
  \bibfield  {author} {\bibinfo {author} {\bibfnamefont {S.}~\bibnamefont
  {Weiss}}\ and\ \bibinfo {author} {\bibfnamefont {G.}~\bibnamefont {Ahlers}},\
  }\bibfield  {title} {\enquote {\bibinfo {title} {Turbulent
  {R}ayleigh--{B}{\'e}nard convection in a cylindrical container with aspect
  ratio $\gamma$= 0.50 and {P}randtl number {P}r= 4.38},}\ }\href {\doibase
  10.1017/S0022112010005963} {\bibfield  {journal} {\bibinfo  {journal}
  {Journal of Fluid Mechanics}\ }\textbf {\bibinfo {volume} {676}},\ \bibinfo
  {pages} {5--40} (\bibinfo {year} {2011})}\BibitemShut {NoStop}%
\bibitem [{\citenamefont {van~der Poel}, \citenamefont {Stevens},\ and\
  \citenamefont {Lohse}(2011)}]{van2011connecting}%
  \BibitemOpen
  \bibfield  {author} {\bibinfo {author} {\bibfnamefont {E.~P.}\ \bibnamefont
  {van~der Poel}}, \bibinfo {author} {\bibfnamefont {R.~J.}\ \bibnamefont
  {Stevens}}, \ and\ \bibinfo {author} {\bibfnamefont {D.}~\bibnamefont
  {Lohse}},\ }\bibfield  {title} {\enquote {\bibinfo {title} {Connecting flow
  structures and heat flux in turbulent {R}ayleigh-{B}{\'e}nard convection},}\
  }\href {\doibase 10.1103/PhysRevE.84.045303} {\bibfield  {journal} {\bibinfo
  {journal} {Physical Review E}\ }\textbf {\bibinfo {volume} {84}},\ \bibinfo
  {pages} {045303} (\bibinfo {year} {2011})}\BibitemShut {NoStop}%
\bibitem [{\citenamefont {van~der Poel}\ \emph {et~al.}(2012)\citenamefont
  {van~der Poel}, \citenamefont {Stevens}, \citenamefont {Sugiyama},\ and\
  \citenamefont {Lohse}}]{van2012flow}%
  \BibitemOpen
  \bibfield  {author} {\bibinfo {author} {\bibfnamefont {E.~P.}\ \bibnamefont
  {van~der Poel}}, \bibinfo {author} {\bibfnamefont {R.~J.}\ \bibnamefont
  {Stevens}}, \bibinfo {author} {\bibfnamefont {K.}~\bibnamefont {Sugiyama}}, \
  and\ \bibinfo {author} {\bibfnamefont {D.}~\bibnamefont {Lohse}},\ }\bibfield
   {title} {\enquote {\bibinfo {title} {Flow states in two-dimensional
  {R}ayleigh-{B}{\'e}nard convection as a function of aspect-ratio and
  {R}ayleigh number},}\ }\href {\doibase 10.1063/1.4744988} {\bibfield
  {journal} {\bibinfo  {journal} {Physics of Fluids}\ }\textbf {\bibinfo
  {volume} {24}},\ \bibinfo {pages} {085104} (\bibinfo {year}
  {2012})}\BibitemShut {NoStop}%
\bibitem [{\citenamefont {Xi}\ \emph {et~al.}(2016)\citenamefont {Xi},
  \citenamefont {Zhang}, \citenamefont {Hao},\ and\ \citenamefont
  {Xia}}]{xi2016higher}%
  \BibitemOpen
  \bibfield  {author} {\bibinfo {author} {\bibfnamefont {H.-D.}\ \bibnamefont
  {Xi}}, \bibinfo {author} {\bibfnamefont {Y.-B.}\ \bibnamefont {Zhang}},
  \bibinfo {author} {\bibfnamefont {J.-T.}\ \bibnamefont {Hao}}, \ and\
  \bibinfo {author} {\bibfnamefont {K.-Q.}\ \bibnamefont {Xia}},\ }\bibfield
  {title} {\enquote {\bibinfo {title} {Higher-order flow modes in turbulent
  {R}ayleigh--{B}{\'e}nard convection},}\ }\href {\doibase
  10.1017/jfm.2016.572} {\bibfield  {journal} {\bibinfo  {journal} {Journal of
  Fluid Mechanics}\ }\textbf {\bibinfo {volume} {805}},\ \bibinfo {pages}
  {31--51} (\bibinfo {year} {2016})}\BibitemShut {NoStop}%
\bibitem [{\citenamefont {P{\'e}tr{\'e}lis}\ \emph {et~al.}(2009)\citenamefont
  {P{\'e}tr{\'e}lis}, \citenamefont {Fauve}, \citenamefont {Dormy},\ and\
  \citenamefont {Valet}}]{petrelis2009simple}%
  \BibitemOpen
  \bibfield  {author} {\bibinfo {author} {\bibfnamefont {F.}~\bibnamefont
  {P{\'e}tr{\'e}lis}}, \bibinfo {author} {\bibfnamefont {S.}~\bibnamefont
  {Fauve}}, \bibinfo {author} {\bibfnamefont {E.}~\bibnamefont {Dormy}}, \ and\
  \bibinfo {author} {\bibfnamefont {J.-P.}\ \bibnamefont {Valet}},\ }\bibfield
  {title} {\enquote {\bibinfo {title} {Simple mechanism for reversals of
  {E}arth's magnetic field},}\ }\href@noop {} {\bibfield  {journal} {\bibinfo
  {journal} {Physical Review Letters}\ }\textbf {\bibinfo {volume} {102}},\
  \bibinfo {pages} {144503} (\bibinfo {year} {2009})}\BibitemShut {NoStop}%
\bibitem [{\citenamefont {Gallet}\ \emph {et~al.}(2012)\citenamefont {Gallet},
  \citenamefont {Herault}, \citenamefont {Laroche}, \citenamefont
  {P{\'e}tr{\'e}lis},\ and\ \citenamefont {Fauve}}]{gallet2012reversals}%
  \BibitemOpen
  \bibfield  {author} {\bibinfo {author} {\bibfnamefont {B.}~\bibnamefont
  {Gallet}}, \bibinfo {author} {\bibfnamefont {J.}~\bibnamefont {Herault}},
  \bibinfo {author} {\bibfnamefont {C.}~\bibnamefont {Laroche}}, \bibinfo
  {author} {\bibfnamefont {F.}~\bibnamefont {P{\'e}tr{\'e}lis}}, \ and\
  \bibinfo {author} {\bibfnamefont {S.}~\bibnamefont {Fauve}},\ }\bibfield
  {title} {\enquote {\bibinfo {title} {Reversals of a large-scale field
  generated over a turbulent background},}\ }\href@noop {} {\bibfield
  {journal} {\bibinfo  {journal} {Geophysical \& Astrophysical Fluid Dynamics}\
  }\textbf {\bibinfo {volume} {106}},\ \bibinfo {pages} {468--492} (\bibinfo
  {year} {2012})}\BibitemShut {NoStop}%
\bibitem [{\citenamefont {Petschel}\ \emph {et~al.}(2011)\citenamefont
  {Petschel}, \citenamefont {Wilczek}, \citenamefont {Breuer}, \citenamefont
  {Friedrich},\ and\ \citenamefont {Hansen}}]{petschel2011statistical}%
  \BibitemOpen
  \bibfield  {author} {\bibinfo {author} {\bibfnamefont {K.}~\bibnamefont
  {Petschel}}, \bibinfo {author} {\bibfnamefont {M.}~\bibnamefont {Wilczek}},
  \bibinfo {author} {\bibfnamefont {M.}~\bibnamefont {Breuer}}, \bibinfo
  {author} {\bibfnamefont {R.}~\bibnamefont {Friedrich}}, \ and\ \bibinfo
  {author} {\bibfnamefont {U.}~\bibnamefont {Hansen}},\ }\bibfield  {title}
  {\enquote {\bibinfo {title} {Statistical analysis of global wind dynamics in
  vigorous {R}ayleigh-{B}{\'e}nard convection},}\ }\href {\doibase
  10.1103/PhysRevE.84.026309} {\bibfield  {journal} {\bibinfo  {journal}
  {Physical Review E}\ }\textbf {\bibinfo {volume} {84}},\ \bibinfo {pages}
  {026309} (\bibinfo {year} {2011})}\BibitemShut {NoStop}%
\bibitem [{\citenamefont {Chandra}\ and\ \citenamefont
  {Verma}(2011)}]{chandra2011dynamics}%
  \BibitemOpen
  \bibfield  {author} {\bibinfo {author} {\bibfnamefont {M.}~\bibnamefont
  {Chandra}}\ and\ \bibinfo {author} {\bibfnamefont {M.~K.}\ \bibnamefont
  {Verma}},\ }\bibfield  {title} {\enquote {\bibinfo {title} {Dynamics and
  symmetries of flow reversals in turbulent convection},}\ }\href {\doibase
  10.1103/PhysRevE.83.067303} {\bibfield  {journal} {\bibinfo  {journal}
  {Physical Review E}\ }\textbf {\bibinfo {volume} {83}},\ \bibinfo {pages}
  {067303} (\bibinfo {year} {2011})}\BibitemShut {NoStop}%
\bibitem [{\citenamefont {Lumley}(1967)}]{lumley1967structure}%
  \BibitemOpen
  \bibfield  {author} {\bibinfo {author} {\bibfnamefont {J.~L.}\ \bibnamefont
  {Lumley}},\ }\href@noop {} {\emph {\bibinfo {title} {The structure of
  inhomogeneous turbulent flows}}}\ (\bibinfo  {publisher} {Nauka},\ \bibinfo
  {year} {1967})\BibitemShut {NoStop}%
\bibitem [{\citenamefont {Berkooz}, \citenamefont {Holmes},\ and\ \citenamefont
  {Lumley}(1993)}]{berkooz1993proper}%
  \BibitemOpen
  \bibfield  {author} {\bibinfo {author} {\bibfnamefont {G.}~\bibnamefont
  {Berkooz}}, \bibinfo {author} {\bibfnamefont {P.}~\bibnamefont {Holmes}}, \
  and\ \bibinfo {author} {\bibfnamefont {J.~L.}\ \bibnamefont {Lumley}},\
  }\bibfield  {title} {\enquote {\bibinfo {title} {The proper orthogonal
  decomposition in the analysis of turbulent flows},}\ }\href {\doibase
  10.1146/annurev.fl.25.010193.002543} {\bibfield  {journal} {\bibinfo
  {journal} {Annual Review of Fluid Mechanics}\ }\textbf {\bibinfo {volume}
  {25}},\ \bibinfo {pages} {539--575} (\bibinfo {year} {1993})}\BibitemShut
  {NoStop}%
\bibitem [{\citenamefont {Chen}\ and\ \citenamefont
  {Doolen}(1998)}]{chen1998lattice}%
  \BibitemOpen
  \bibfield  {author} {\bibinfo {author} {\bibfnamefont {S.}~\bibnamefont
  {Chen}}\ and\ \bibinfo {author} {\bibfnamefont {G.~D.}\ \bibnamefont
  {Doolen}},\ }\bibfield  {title} {\enquote {\bibinfo {title} {Lattice
  {B}oltzmann method for fluid flows},}\ }\href {\doibase
  10.1146/annurev.fluid.30.1.329} {\bibfield  {journal} {\bibinfo  {journal}
  {Annual Review of Fluid Mechanics}\ }\textbf {\bibinfo {volume} {30}},\
  \bibinfo {pages} {329--364} (\bibinfo {year} {1998})}\BibitemShut {NoStop}%
\bibitem [{\citenamefont {Aidun}\ and\ \citenamefont
  {Clausen}(2010)}]{aidun2010lattice}%
  \BibitemOpen
  \bibfield  {author} {\bibinfo {author} {\bibfnamefont {C.~K.}\ \bibnamefont
  {Aidun}}\ and\ \bibinfo {author} {\bibfnamefont {J.~R.}\ \bibnamefont
  {Clausen}},\ }\bibfield  {title} {\enquote {\bibinfo {title}
  {Lattice-{B}oltzmann method for complex flows},}\ }\href {\doibase
  10.1146/annurev-fluid-121108-145519} {\bibfield  {journal} {\bibinfo
  {journal} {Annual Review of Fluid Mechanics}\ }\textbf {\bibinfo {volume}
  {42}},\ \bibinfo {pages} {439--472} (\bibinfo {year} {2010})}\BibitemShut
  {NoStop}%
\bibitem [{\citenamefont {Xu}, \citenamefont {Shyy},\ and\ \citenamefont
  {Zhao}(2017)}]{xu2017lattice}%
  \BibitemOpen
  \bibfield  {author} {\bibinfo {author} {\bibfnamefont {A.}~\bibnamefont
  {Xu}}, \bibinfo {author} {\bibfnamefont {W.}~\bibnamefont {Shyy}}, \ and\
  \bibinfo {author} {\bibfnamefont {T.}~\bibnamefont {Zhao}},\ }\bibfield
  {title} {\enquote {\bibinfo {title} {Lattice {B}oltzmann modeling of
  transport phenomena in fuel cells and flow batteries},}\ }\href {\doibase
  10.1007/s10409-017-0667-6} {\bibfield  {journal} {\bibinfo  {journal} {Acta
  Mechanica Sinica}\ }\textbf {\bibinfo {volume} {33}},\ \bibinfo {pages}
  {555--574} (\bibinfo {year} {2017})}\BibitemShut {NoStop}%
\bibitem [{\citenamefont {Huang}, \citenamefont {Sukop},\ and\ \citenamefont
  {Lu}(2015)}]{huang2015multiphase}%
  \BibitemOpen
  \bibfield  {author} {\bibinfo {author} {\bibfnamefont {H.}~\bibnamefont
  {Huang}}, \bibinfo {author} {\bibfnamefont {M.}~\bibnamefont {Sukop}}, \ and\
  \bibinfo {author} {\bibfnamefont {X.}~\bibnamefont {Lu}},\ }\href@noop {}
  {\emph {\bibinfo {title} {Multiphase lattice Boltzmann methods: {T}heory and
  application}}}\ (\bibinfo  {publisher} {John Wiley \& Sons},\ \bibinfo {year}
  {2015})\BibitemShut {NoStop}%
\bibitem [{\citenamefont {Xu}, \citenamefont {Shi},\ and\ \citenamefont
  {Zhao}(2017)}]{xu2017accelerated}%
  \BibitemOpen
  \bibfield  {author} {\bibinfo {author} {\bibfnamefont {A.}~\bibnamefont
  {Xu}}, \bibinfo {author} {\bibfnamefont {L.}~\bibnamefont {Shi}}, \ and\
  \bibinfo {author} {\bibfnamefont {T.}~\bibnamefont {Zhao}},\ }\bibfield
  {title} {\enquote {\bibinfo {title} {Accelerated lattice {B}oltzmann
  simulation using {GPU} and {O}pen{ACC} with data management},}\ }\href
  {\doibase 10.1016/j.ijheatmasstransfer.2017.02.032} {\bibfield  {journal}
  {\bibinfo  {journal} {International Journal of Heat and Mass Transfer}\
  }\textbf {\bibinfo {volume} {109}},\ \bibinfo {pages} {577--588} (\bibinfo
  {year} {2017})}\BibitemShut {NoStop}%
\bibitem [{\citenamefont {Xu}, \citenamefont {Shi},\ and\ \citenamefont
  {Xi}(2019{\natexlab{a}})}]{xu2019lattice}%
  \BibitemOpen
  \bibfield  {author} {\bibinfo {author} {\bibfnamefont {A.}~\bibnamefont
  {Xu}}, \bibinfo {author} {\bibfnamefont {L.}~\bibnamefont {Shi}}, \ and\
  \bibinfo {author} {\bibfnamefont {H.-D.}\ \bibnamefont {Xi}},\ }\bibfield
  {title} {\enquote {\bibinfo {title} {Lattice {B}oltzmann simulations of
  three-dimensional thermal convective flows at high {R}ayleigh number},}\
  }\href {\doibase 10.1016/j.ijheatmasstransfer.2019.06.002} {\bibfield
  {journal} {\bibinfo  {journal} {International Journal of Heat and Mass
  Transfer}\ }\textbf {\bibinfo {volume} {140}},\ \bibinfo {pages} {359--370}
  (\bibinfo {year} {2019}{\natexlab{a}})}\BibitemShut {NoStop}%
\bibitem [{\citenamefont {Xu}, \citenamefont {Shi},\ and\ \citenamefont
  {Xi}(2019{\natexlab{b}})}]{xu2019statistics}%
  \BibitemOpen
  \bibfield  {author} {\bibinfo {author} {\bibfnamefont {A.}~\bibnamefont
  {Xu}}, \bibinfo {author} {\bibfnamefont {L.}~\bibnamefont {Shi}}, \ and\
  \bibinfo {author} {\bibfnamefont {H.-D.}\ \bibnamefont {Xi}},\ }\bibfield
  {title} {\enquote {\bibinfo {title} {Statistics of temperature and thermal
  energy dissipation rate in low-{P}randtl number turbulent thermal
  convection},}\ }\href {\doibase 10.1063/1.5129818} {\bibfield  {journal}
  {\bibinfo  {journal} {Physics of Fluids}\ }\textbf {\bibinfo {volume} {31}},\
  \bibinfo {pages} {125101} (\bibinfo {year} {2019}{\natexlab{b}})}\BibitemShut
  {NoStop}%
\bibitem [{\citenamefont {Xu}\ \emph {et~al.}(2020)\citenamefont {Xu},
  \citenamefont {Tao}, \citenamefont {Shi},\ and\ \citenamefont
  {Xi}}]{xu2020transport}%
  \BibitemOpen
  \bibfield  {author} {\bibinfo {author} {\bibfnamefont {A.}~\bibnamefont
  {Xu}}, \bibinfo {author} {\bibfnamefont {S.}~\bibnamefont {Tao}}, \bibinfo
  {author} {\bibfnamefont {L.}~\bibnamefont {Shi}}, \ and\ \bibinfo {author}
  {\bibfnamefont {H.-D.}\ \bibnamefont {Xi}},\ }\bibfield  {title} {\enquote
  {\bibinfo {title} {Transport and deposition of dilute microparticles in
  turbulent thermal convection},}\ }\href {\doibase 10.1063/5.0018804}
  {\bibfield  {journal} {\bibinfo  {journal} {Physics of Fluids}\ }\textbf
  {\bibinfo {volume} {32}},\ \bibinfo {pages} {083301} (\bibinfo {year}
  {2020})}\BibitemShut {NoStop}%
\bibitem [{\citenamefont {Kolmogorov}(1941)}]{kolmogorov1941local}%
  \BibitemOpen
  \bibfield  {author} {\bibinfo {author} {\bibfnamefont {A.~N.}\ \bibnamefont
  {Kolmogorov}},\ }\bibfield  {title} {\enquote {\bibinfo {title} {The local
  structure of turbulence in incompressible viscous fluid for very large
  {R}eynolds numbers},}\ }\href@noop {} {\bibfield  {journal} {\bibinfo
  {journal} {Doklady Akademii Nauk SSSR}\ }\textbf {\bibinfo {volume} {30}},\
  \bibinfo {pages} {301--305} (\bibinfo {year} {1941})}\BibitemShut {NoStop}%
\bibitem [{\citenamefont {Batchelor}(1959)}]{batchelor1959small}%
  \BibitemOpen
  \bibfield  {author} {\bibinfo {author} {\bibfnamefont {G.~K.}\ \bibnamefont
  {Batchelor}},\ }\bibfield  {title} {\enquote {\bibinfo {title} {Small-scale
  variation of convected quantities like temperature in turbulent fluid {P}art
  1. {G}eneral discussion and the case of small conductivity},}\ }\href@noop {}
  {\bibfield  {journal} {\bibinfo  {journal} {Journal of Fluid Mechanics}\
  }\textbf {\bibinfo {volume} {5}},\ \bibinfo {pages} {113--133} (\bibinfo
  {year} {1959})}\BibitemShut {NoStop}%
\bibitem [{\citenamefont {Silano}, \citenamefont {Sreenivasan},\ and\
  \citenamefont {Verzicco}(2010)}]{silano2010numerical}%
  \BibitemOpen
  \bibfield  {author} {\bibinfo {author} {\bibfnamefont {G.}~\bibnamefont
  {Silano}}, \bibinfo {author} {\bibfnamefont {K.}~\bibnamefont {Sreenivasan}},
  \ and\ \bibinfo {author} {\bibfnamefont {R.}~\bibnamefont {Verzicco}},\
  }\bibfield  {title} {\enquote {\bibinfo {title} {Numerical simulations of
  {R}ayleigh--{B}{\'e}nard convection for {P}randtl numbers between $10^{-1}$
  and $10^{4}$ and rayleigh numbers between $10^{5}$ and $10^{9}$},}\ }\href
  {\doibase 10.1017/S0022112010003290} {\bibfield  {journal} {\bibinfo
  {journal} {Journal of Fluid Mechanics}\ }\textbf {\bibinfo {volume} {662}},\
  \bibinfo {pages} {409--446} (\bibinfo {year} {2010})}\BibitemShut {NoStop}%
\bibitem [{\citenamefont {Auma{\^\i}tre}\ and\ \citenamefont
  {Fauve}(2003)}]{aumaitre2003statistical}%
  \BibitemOpen
  \bibfield  {author} {\bibinfo {author} {\bibfnamefont {S.}~\bibnamefont
  {Auma{\^\i}tre}}\ and\ \bibinfo {author} {\bibfnamefont {S.}~\bibnamefont
  {Fauve}},\ }\bibfield  {title} {\enquote {\bibinfo {title} {Statistical
  properties of the fluctuations of the heat transfer in turbulent
  convection},}\ }\href@noop {} {\bibfield  {journal} {\bibinfo  {journal} {EPL
  (Europhysics Letters)}\ }\textbf {\bibinfo {volume} {62}},\ \bibinfo {pages}
  {822} (\bibinfo {year} {2003})}\BibitemShut {NoStop}%
\bibitem [{\citenamefont {Zhang}, \citenamefont {Zhou},\ and\ \citenamefont
  {Sun}(2017)}]{zhang2017statisticsJFM}%
  \BibitemOpen
  \bibfield  {author} {\bibinfo {author} {\bibfnamefont {Y.}~\bibnamefont
  {Zhang}}, \bibinfo {author} {\bibfnamefont {Q.}~\bibnamefont {Zhou}}, \ and\
  \bibinfo {author} {\bibfnamefont {C.}~\bibnamefont {Sun}},\ }\bibfield
  {title} {\enquote {\bibinfo {title} {Statistics of kinetic and thermal energy
  dissipation rates in two-dimensional turbulent {R}ayleigh--{B}{\'e}nard
  convection},}\ }\href {\doibase 10.1017/jfm.2017.19} {\bibfield  {journal}
  {\bibinfo  {journal} {Journal of Fluid Mechanics}\ }\textbf {\bibinfo
  {volume} {814}},\ \bibinfo {pages} {165--184} (\bibinfo {year}
  {2017})}\BibitemShut {NoStop}%
\bibitem [{\citenamefont {Chandra}\ and\ \citenamefont
  {Verma}(2013)}]{chandra2013flow}%
  \BibitemOpen
  \bibfield  {author} {\bibinfo {author} {\bibfnamefont {M.}~\bibnamefont
  {Chandra}}\ and\ \bibinfo {author} {\bibfnamefont {M.~K.}\ \bibnamefont
  {Verma}},\ }\bibfield  {title} {\enquote {\bibinfo {title} {Flow reversals in
  turbulent convection via vortex reconnections},}\ }\href {\doibase
  10.1103/PhysRevLett.110.114503} {\bibfield  {journal} {\bibinfo  {journal}
  {Physical Review Letters}\ }\textbf {\bibinfo {volume} {110}},\ \bibinfo
  {pages} {114503} (\bibinfo {year} {2013})}\BibitemShut {NoStop}%
\bibitem [{\citenamefont {Verma}, \citenamefont {Ambhire},\ and\ \citenamefont
  {Pandey}(2015)}]{verma2015flow}%
  \BibitemOpen
  \bibfield  {author} {\bibinfo {author} {\bibfnamefont {M.~K.}\ \bibnamefont
  {Verma}}, \bibinfo {author} {\bibfnamefont {S.~C.}\ \bibnamefont {Ambhire}},
  \ and\ \bibinfo {author} {\bibfnamefont {A.}~\bibnamefont {Pandey}},\
  }\bibfield  {title} {\enquote {\bibinfo {title} {Flow reversals in turbulent
  convection with free-slip walls},}\ }\href {\doibase 10.1063/1.4918590}
  {\bibfield  {journal} {\bibinfo  {journal} {Physics of Fluids}\ }\textbf
  {\bibinfo {volume} {27}},\ \bibinfo {pages} {047102} (\bibinfo {year}
  {2015})}\BibitemShut {NoStop}%
\bibitem [{\citenamefont {Wang}\ \emph {et~al.}(2018)\citenamefont {Wang},
  \citenamefont {Xia}, \citenamefont {Wang}, \citenamefont {Sun}, \citenamefont
  {Zhou},\ and\ \citenamefont {Wan}}]{wang2018flow}%
  \BibitemOpen
  \bibfield  {author} {\bibinfo {author} {\bibfnamefont {Q.}~\bibnamefont
  {Wang}}, \bibinfo {author} {\bibfnamefont {S.-N.}\ \bibnamefont {Xia}},
  \bibinfo {author} {\bibfnamefont {B.-F.}\ \bibnamefont {Wang}}, \bibinfo
  {author} {\bibfnamefont {D.-J.}\ \bibnamefont {Sun}}, \bibinfo {author}
  {\bibfnamefont {Q.}~\bibnamefont {Zhou}}, \ and\ \bibinfo {author}
  {\bibfnamefont {Z.-H.}\ \bibnamefont {Wan}},\ }\bibfield  {title} {\enquote
  {\bibinfo {title} {Flow reversals in two-dimensional thermal convection in
  tilted cells},}\ }\href {\doibase 10.1017/jfm.2018.451} {\bibfield  {journal}
  {\bibinfo  {journal} {Journal of Fluid Mechanics}\ }\textbf {\bibinfo
  {volume} {849}},\ \bibinfo {pages} {355--372} (\bibinfo {year}
  {2018})}\BibitemShut {NoStop}%
\bibitem [{\citenamefont {Chen}\ \emph {et~al.}(2019)\citenamefont {Chen},
  \citenamefont {Huang}, \citenamefont {Xia},\ and\ \citenamefont
  {Xi}}]{chen2019emergence}%
  \BibitemOpen
  \bibfield  {author} {\bibinfo {author} {\bibfnamefont {X.}~\bibnamefont
  {Chen}}, \bibinfo {author} {\bibfnamefont {S.-D.}\ \bibnamefont {Huang}},
  \bibinfo {author} {\bibfnamefont {K.-Q.}\ \bibnamefont {Xia}}, \ and\
  \bibinfo {author} {\bibfnamefont {H.-D.}\ \bibnamefont {Xi}},\ }\bibfield
  {title} {\enquote {\bibinfo {title} {Emergence of substructures inside the
  large-scale circulation induces transition in flow reversals in turbulent
  thermal convection},}\ }\href@noop {} {\bibfield  {journal} {\bibinfo
  {journal} {Journal of Fluid Mechanics}\ }\textbf {\bibinfo {volume} {877}}
  (\bibinfo {year} {2019})}\BibitemShut {NoStop}%
\bibitem [{\citenamefont {Wagner}\ and\ \citenamefont
  {Shishkina}(2013)}]{wagner2013aspect}%
  \BibitemOpen
  \bibfield  {author} {\bibinfo {author} {\bibfnamefont {S.}~\bibnamefont
  {Wagner}}\ and\ \bibinfo {author} {\bibfnamefont {O.}~\bibnamefont
  {Shishkina}},\ }\bibfield  {title} {\enquote {\bibinfo {title} {Aspect-ratio
  dependency of {R}ayleigh-{B}{\'e}nard convection in box-shaped containers},}\
  }\href {\doibase 10.1063/1.4819141} {\bibfield  {journal} {\bibinfo
  {journal} {Physics of Fluids}\ }\textbf {\bibinfo {volume} {25}},\ \bibinfo
  {pages} {085110} (\bibinfo {year} {2013})}\BibitemShut {NoStop}%
\bibitem [{\citenamefont {Chong}\ \emph {et~al.}(2018)\citenamefont {Chong},
  \citenamefont {Wagner}, \citenamefont {Kaczorowski}, \citenamefont
  {Shishkina},\ and\ \citenamefont {Xia}}]{chong2018effect}%
  \BibitemOpen
  \bibfield  {author} {\bibinfo {author} {\bibfnamefont {K.~L.}\ \bibnamefont
  {Chong}}, \bibinfo {author} {\bibfnamefont {S.}~\bibnamefont {Wagner}},
  \bibinfo {author} {\bibfnamefont {M.}~\bibnamefont {Kaczorowski}}, \bibinfo
  {author} {\bibfnamefont {O.}~\bibnamefont {Shishkina}}, \ and\ \bibinfo
  {author} {\bibfnamefont {K.-Q.}\ \bibnamefont {Xia}},\ }\bibfield  {title}
  {\enquote {\bibinfo {title} {Effect of {P}randtl number on heat transport
  enhancement in {R}ayleigh-{B}{\'e}nard convection under geometrical
  confinement},}\ }\href {\doibase 10.1103/PhysRevFluids.3.013501} {\bibfield
  {journal} {\bibinfo  {journal} {Physical Review Fluids}\ }\textbf {\bibinfo
  {volume} {3}},\ \bibinfo {pages} {013501} (\bibinfo {year}
  {2018})}\BibitemShut {NoStop}%
\bibitem [{\citenamefont {Xia}, \citenamefont {Sun},\ and\ \citenamefont
  {Zhou}(2003)}]{xia2003particle}%
  \BibitemOpen
  \bibfield  {author} {\bibinfo {author} {\bibfnamefont {K.-Q.}\ \bibnamefont
  {Xia}}, \bibinfo {author} {\bibfnamefont {C.}~\bibnamefont {Sun}}, \ and\
  \bibinfo {author} {\bibfnamefont {S.-Q.}\ \bibnamefont {Zhou}},\ }\bibfield
  {title} {\enquote {\bibinfo {title} {Particle image velocimetry measurement
  of the velocity field in turbulent thermal convection},}\ }\href@noop {}
  {\bibfield  {journal} {\bibinfo  {journal} {Physical review E}\ }\textbf
  {\bibinfo {volume} {68}},\ \bibinfo {pages} {066303} (\bibinfo {year}
  {2003})}\BibitemShut {NoStop}%
\bibitem [{\citenamefont {Podvin}\ and\ \citenamefont
  {Sergent}(2015)}]{podvin2015large}%
  \BibitemOpen
  \bibfield  {author} {\bibinfo {author} {\bibfnamefont {B.}~\bibnamefont
  {Podvin}}\ and\ \bibinfo {author} {\bibfnamefont {A.}~\bibnamefont
  {Sergent}},\ }\bibfield  {title} {\enquote {\bibinfo {title} {A large-scale
  investigation of wind reversal in a square {R}ayleigh--{B}{\'e}nard cell},}\
  }\href {\doibase 10.1017/jfm.2015.15} {\bibfield  {journal} {\bibinfo
  {journal} {Journal of Fluid Mechanics}\ }\textbf {\bibinfo {volume} {766}},\
  \bibinfo {pages} {172--201} (\bibinfo {year} {2015})}\BibitemShut {NoStop}%
\bibitem [{\citenamefont {Podvin}\ and\ \citenamefont
  {Sergent}(2017)}]{podvin2017precursor}%
  \BibitemOpen
  \bibfield  {author} {\bibinfo {author} {\bibfnamefont {B.}~\bibnamefont
  {Podvin}}\ and\ \bibinfo {author} {\bibfnamefont {A.}~\bibnamefont
  {Sergent}},\ }\bibfield  {title} {\enquote {\bibinfo {title} {Precursor for
  wind reversal in a square {R}ayleigh-{B}{\'e}nard cell},}\ }\href {\doibase
  10.1103/PhysRevE.95.013112} {\bibfield  {journal} {\bibinfo  {journal}
  {Physical Review E}\ }\textbf {\bibinfo {volume} {95}},\ \bibinfo {pages}
  {013112} (\bibinfo {year} {2017})}\BibitemShut {NoStop}%
\bibitem [{\citenamefont {Castillo-Castellanos}\ \emph
  {et~al.}(2019)\citenamefont {Castillo-Castellanos}, \citenamefont {Sergent},
  \citenamefont {Podvin},\ and\ \citenamefont {Rossi}}]{castillo2019cessation}%
  \BibitemOpen
  \bibfield  {author} {\bibinfo {author} {\bibfnamefont {A.}~\bibnamefont
  {Castillo-Castellanos}}, \bibinfo {author} {\bibfnamefont {A.}~\bibnamefont
  {Sergent}}, \bibinfo {author} {\bibfnamefont {B.}~\bibnamefont {Podvin}}, \
  and\ \bibinfo {author} {\bibfnamefont {M.}~\bibnamefont {Rossi}},\ }\bibfield
   {title} {\enquote {\bibinfo {title} {Cessation and reversals of large-scale
  structures in square {R}ayleigh--{B}{\'e}nard cells},}\ }\href {\doibase
  10.1017/jfm.2019.598} {\bibfield  {journal} {\bibinfo  {journal} {Journal of
  Fluid Mechanics}\ }\textbf {\bibinfo {volume} {877}},\ \bibinfo {pages}
  {922--954} (\bibinfo {year} {2019})}\BibitemShut {NoStop}%
\bibitem [{\citenamefont {Soucasse}\ \emph {et~al.}(2019)\citenamefont
  {Soucasse}, \citenamefont {Podvin}, \citenamefont {Rivi{\`e}re},\ and\
  \citenamefont {Soufiani}}]{soucasse2019proper}%
  \BibitemOpen
  \bibfield  {author} {\bibinfo {author} {\bibfnamefont {L.}~\bibnamefont
  {Soucasse}}, \bibinfo {author} {\bibfnamefont {B.}~\bibnamefont {Podvin}},
  \bibinfo {author} {\bibfnamefont {P.}~\bibnamefont {Rivi{\`e}re}}, \ and\
  \bibinfo {author} {\bibfnamefont {A.}~\bibnamefont {Soufiani}},\ }\bibfield
  {title} {\enquote {\bibinfo {title} {Proper orthogonal decomposition analysis
  and modelling of large-scale flow reorientations in a cubic
  {R}ayleigh--{B}{\'e}nard cell},}\ }\href {\doibase 10.1017/jfm.2019.746}
  {\bibfield  {journal} {\bibinfo  {journal} {Journal of Fluid Mechanics}\
  }\textbf {\bibinfo {volume} {881}},\ \bibinfo {pages} {23--50} (\bibinfo
  {year} {2019})}\BibitemShut {NoStop}%
\bibitem [{\citenamefont {Kutz}\ \emph {et~al.}(2016)\citenamefont {Kutz},
  \citenamefont {Brunton}, \citenamefont {Brunton},\ and\ \citenamefont
  {Proctor}}]{kutz2016dynamic}%
  \BibitemOpen
  \bibfield  {author} {\bibinfo {author} {\bibfnamefont {J.~N.}\ \bibnamefont
  {Kutz}}, \bibinfo {author} {\bibfnamefont {S.~L.}\ \bibnamefont {Brunton}},
  \bibinfo {author} {\bibfnamefont {B.~W.}\ \bibnamefont {Brunton}}, \ and\
  \bibinfo {author} {\bibfnamefont {J.~L.}\ \bibnamefont {Proctor}},\
  }\href@noop {} {\emph {\bibinfo {title} {Dynamic mode decomposition:
  data-driven modeling of complex systems}}}\ (\bibinfo  {publisher} {SIAM},\
  \bibinfo {year} {2016})\BibitemShut {NoStop}%
\bibitem [{\citenamefont {Brunton}\ and\ \citenamefont
  {Kutz}(2019)}]{brunton2019data}%
  \BibitemOpen
  \bibfield  {author} {\bibinfo {author} {\bibfnamefont {S.~L.}\ \bibnamefont
  {Brunton}}\ and\ \bibinfo {author} {\bibfnamefont {J.~N.}\ \bibnamefont
  {Kutz}},\ }\href@noop {} {\emph {\bibinfo {title} {Data-driven science and
  engineering: {M}achine learning, dynamical systems, and control}}}\ (\bibinfo
   {publisher} {Cambridge University Press},\ \bibinfo {year}
  {2019})\BibitemShut {NoStop}%
\end{thebibliography}%

\end{document}